%% file: AliRpPbHFM.tex
\documentclass[ALICE,manyauthors]{cernphprep}

\usepackage[comma,square,numbers,sort&compress]{natbib}
\usepackage{hyperref}
\usepackage{lineno}
\usepackage{multirow}
\usepackage[utf8]{inputenc}
\usepackage{color}

\newcommand{\pt}{$p_{\mathrm T}$}

\DeclareUnicodeCharacter{00A0}{ }

\begin{document}%

\begin{titlepage}
\PHyear{2017}
\PHnumber{022}      
\PHdate{31 January}  
%

\title{Production of muons from heavy--flavour hadron decays \\ 
in p--Pb collisions at $\mathbf{\sqrt{{\textit s}_{NN}}  = 5.02}$ TeV}
\ShortTitle{Heavy-flavour decay muon production in p--Pb collisions} 

\Collaboration{ALICE Collaboration\thanks{See Appendix~\ref{app:collab} for 
the list of collaboration members}}
\ShortAuthor{ALICE Collaboration} 

\begin{abstract}
\input{c00Abstract.tex}
\end{abstract}
\end{titlepage}
\setcounter{page}{2}

\input{c01Introduction.tex}
\input{c02DataSample.tex}
\input{c03Analysis.tex}
\input{c04Results.tex}
\input{c05Conclusion.tex}

\newenvironment{acknowledgement}{\relax}{\relax}
\begin{acknowledgement}
\section{Acknowledgements}
\input{fa_2017-01-25.tex}
\end{acknowledgement}

\bibliographystyle{utphys}
\bibliography{AliRpPbHFM}
\newpage
%
%
\appendix
\section{The ALICE Collaboration}
\label{app:collab}
\input{Alice_Authorlist_2017-Jan-25_mod.tex}  
\end{document}

%% file: c00Abstract.tex
The production of muons from heavy-flavour hadron decays in p--Pb collisions 
at $\sqrt{{\textit s}_{\rm NN}}=5.02~{\rm TeV}$ was studied 
for $2 < p_{\rm T} < 16$~GeV/$c$ with the ALICE detector at the CERN LHC. 
The measurement was performed
at forward (p-going direction) and backward (Pb-going direction) rapidity, 
in the ranges of rapidity in 
the centre-of-mass system (cms)
$2.03<y_{\rm cms}<3.53$ and $-4.46<y_{\rm cms}<-2.96$,
respectively. The production cross sections and nuclear modification
factors are presented as a function of transverse momentum (\pt).
At forward rapidity, the nuclear 
modification factor is compatible with unity while at backward rapidity, in 
the interval $2.5<p_{\rm T}<3.5$~GeV/$c$,
it is above unity by more than 2$\sigma$.
The ratio of the forward-to-backward
production cross sections is also measured in the overlapping interval 
$2.96 < \vert y_{\rm cms} \vert < 3.53$ and is smaller than 
unity by 3.7$\sigma$ in $2.5<p_{\rm T}<3.5$~GeV/$c$. 
The data are described by 
model calculations including 
cold nuclear matter effects.

%% file: c01Introduction.tex
\section{Introduction}\label{sec:intro}

The study of ultra-relativistic heavy-ion collisions aims at investigating the
properties of strongly-in\-ter\-acting matter under extreme conditions of
temperature and energy density. Under these conditions,
Quantum Chromodynamics (QCD) calculations on the lattice
predict a transition to 
a Quark-Gluon Plasma (QGP) in which colour confinement vanishes and 
chiral symmetry is partially restored~\cite{Karsch:2006xs, Bazavov:2011nk}. 
Heavy quarks (charm and beauty) are essential probes of the 
properties of the QGP since they are produced in hard 
scattering processes in the early stage of the collision and, while 
propagating 
through the medium, interact with the QGP constituents.
The nuclear modification factor $R_{\rm AA}$ 
is commonly used to characterise heavy-quark interaction with the medium
constituents. 
It is defined as 
the ratio between the particle yield in nucleus-nucleus (AA) collisions 
and a reference obtained by scaling the yield measured in proton-proton (pp) 
collisions
by the number of binary nucleon-nucleon collisions, calculated
with the Glauber model~\cite{Miller:2007ri}. 
 Heavy-quark production in pp collisions 
at various energies is described within uncertainties 
by perturbative QCD (pQCD) 
calculations~\cite{Abelev:2012pi, Abelev:2012qh, Abelev:2012xe, Abelev:2012tca,
Abelev:2012vra, Abelev:2012sca, Abelev:2014hla, Abelev:2014gla}. 
In central Pb--Pb collisions ($\sqrt{s_{\rm NN}} = 2.76$~TeV), a suppression  
of D mesons and leptons from heavy-flavour hadron decays by a factor of 
about 3--5 was measured for transverse momenta 
$p_{\rm T} > 4$~GeV/$c$~\cite{ALICE:2012ab,Abelev:2012qh,Adam:2015sza,
Adam:2016khe}.
Further insights into the QGP evolution and the in-medium interactions 
can be gained from the study of the particle azimuthal anisotropy expressed in 
terms of Fourier series, where the second order coefficient $v_2$ is the 
elliptic flow. A positive $v_2$ was observed at low 
and/or intermediate $p_{\rm T}$ in semi-central 
Pb--Pb collisions for D mesons and electrons from heavy-flavour hadron decays 
at mid-rapidity~\cite{Abelev:2013lca,Abelev:2014ipa,Adam:2016ssk} 
and for muons from heavy-flavour hadron decays at forward 
rapidity~\cite{Adam:2015pga}, 
confirming the significant interaction of heavy quarks with the medium 
constituents.

Although the suppression of high-\pt~particle yield suggests 
that heavy quarks lose a significant amount of their initial 
energy~\cite{He:2014cla,Nahrgang:2013xaa,Uphoff:2012gb,Wicks:2005gt,
Horowitz:2011wm,Lang:2012cx,Cao:2013ita}, this 
suppression cannot be, a priori, exclusively attributed 
to the interaction of quarks with the hot and dense medium formed in 
the collision. 
Indeed, for a comprehensive understanding of Pb--Pb results, it is  
fundamental to 
quantify Cold Nuclear Matter (CNM) 
effects, which can modify the \pt~spectra in nuclear 
collisions independently from the formation of a QGP.
Cold nuclear matter effects include the modification of the Parton 
Distribution Functions (PDFs) of the nuclei with respect 
to a superposition of nucleon PDFs, addressed by 
nuclear shadowing models~\cite{Eskola:2009uj,Helenius:2012wd} or 
gluon saturation 
models as the Colour Glass Condensate (CGC) effective 
theory~\cite{Fujii:2013yja,Albacete:2012xq}. Other CNM effects are 
Cronin enhancement through 
$k_{\rm T}$~broadening~\cite{Lev:1983hh,Wang:1998ww,Kopeliovich:2002yh} and
energy loss in the initial~\cite{Vitev:2007ve} and final stages 
of the collision. These effects can be assessed by studying p--Pb collisions, 
where the formation of an extended hot and dense system is not expected. 
A possible presence of final-state effects in small systems 
at RHIC and LHC energies is suggested by measurements of long-range 
correlations~\cite{Aad:2012gla,ABELEV:2013wsa,Abelev:2012ola,CMS:2012qk,
Adam:2015bka} consistent with the presence of collective effects. 
This is further supported by the measurements of the 
species-dependent nuclear modification factors of identified particles 
in d--Au collisions~\cite{Adler:2006xd}, multiplicity dependence of 
$\rm \pi^\pm$, $\rm K^\pm$, $\rm p$ and $\Lambda$ production in p--Pb 
collisions~\cite{Abelev:2013haa}, 
and a significant suppression of $\psi$(2S) yields in comparison to those
of $\rm J/\psi$~\cite{Abelev:2014zpa,Adare:2013ezl}.

Cold nuclear matter effects on heavy-flavour production have been 
thoroughly investigated at RHIC by 
the PHENIX and STAR Collaborations through the measurement of the 
production of 
leptons from heavy-flavour hadron decays in d--Au collisions at 
$\sqrt{s_{\rm NN}}$~=~200~GeV. 
An enhancement of the yields of electrons from heavy-flavour 
hadron decays, 
with respect to a binary-scaled pp reference, was 
observed at mid-rapidity~\cite{Adare:2012yxa, Abelev:2006db}.   
An enhancement (suppression) 
of muons from heavy-flavour hadron decays was measured at backward (forward) 
rapidity~\cite{Adare:2013lkk}. The differences observed 
between forward and backward 
rapidity are not reproduced by models based only on 
modifications of the initial parton densities~\cite{Helenius:2012wd}.
Finally, the recent measurement of 
azimuthal correlations between electrons from heavy-flavour hadron decays at 
mid-rapidity and muons from heavy-flavour hadron decays at forward 
rapidity~\cite{Adare:2013xlp} shows a 
suppression of the yield of electron-muon pairs with $\Delta\phi =\pi$,  
suggesting that CNM effects modify the c$\rm \bar c$ 
correlations. An experimental effort to 
quantify CNM effects on heavy-flavour production is underway also at the LHC. 
The measurement of the \pt-integrated nuclear modification factor of 
$\rm J/\psi$ from B-hadron 
decays in p--Pb collisions at $\sqrt{s_{\rm NN}}$ = 5.02 TeV by 
the LHCb Collaboration~\cite{Aaij:2013zxa}
indicates a suppression by about 20\% 
at forward rapidity and no suppression at backward rapidity. 
The measurements of the nuclear modification factors of 
B$^+$, B$^0$ and B$_{\rm s}^0$ by the CMS 
Collaboration~\cite{Khachatryan:2015uja} 
and of the forward-to-backward ratio of $\rm J/\psi$ from B-hadron decays by 
the ATLAS Collaboration~\cite{Aad:2015ddl} at high \pt~are also compatible 
with unity. The mid-rapidity nuclear modification factors of prompt 
D mesons~\cite{Abelev:2014hha} and electrons 
from heavy-flavour hadron and beauty-hadron 
decays~\cite{Adam:2015qda,Adam:2016wyz} 
measured by the ALICE Collaboration are found consistent with unity. 

This Letter presents differential measurements of the production of muons from 
heavy-flavour hadron decays 
for $2 < p_{\rm T} < 16$~GeV/$c$ in p--Pb collisions 
at $\sqrt {s_{\rm NN}}$ = 5.02 TeV at forward and backward rapidity 
performed by the ALICE Collaboration at the LHC. 
Comparisons with model calculations to extract 
relevant information concerning CNM effects are also discussed. These 
measurements 
cover forward ($2.03<y_{\rm cms}<3.53$, p-going direction) and backward 
($-4.46<y_{\rm cms}<-2.96$, Pb-going direction) rapidity regions. 
The Bjorken-$x$ values
of gluons in the Pb nucleus probed by 
measurements of muons from heavy-flavour hadron decays have been 
estimated with PYTHIA 8 (Tune 4C)~\cite{Sjostrand:2014zea} considering 
Leading Order (pair creation) and Next-to-Leading Order 
(flavour excitation and gluon splitting) processes. At forward rapidity, 
they are located
in the range from about $5 \cdot 10^{-6}$ to $10^{-2}$ and the median of 
the distribution is about 10$^{-4}$. 
At backward rapidity, the 
Bjorken-$x$ values are expected to vary from about $10^{-3}$ to $10^{-1}$ and 
the median is of the order of $10^{-2}$.

The Letter is structured as follows. Section~2 
describes the apparatus with an emphasis on the detectors used in the 
analysis and the data taking 
conditions. Section~3 addresses the analysis 
details. 
Section 4 presents the 
results, namely the \pt-differential cross sections 
and nuclear modification factors at forward 
and backward rapidity and the
forward-to-backward ratio in a smaller overlapping rapidity interval 
($2.96 < \vert y_{\rm cms} \vert <3.53$). 
Finally, the results are compared with model calculations which include 
CNM effects.

%% file: c02DataSample.tex
\section{Experimental apparatus and data samples}\label{sec:det}

A detailed description of the ALICE detector is available 
in~\cite{Aamodt:2008zz} and its performance is discussed 
in~\cite{Abelev:2014ffa}. 
Muons are detected in ALICE using the muon spectrometer in the 
pseudo-rapidity interval $-4 < \eta_{\rm lab} < -2.5$. The 
muon spectrometer consists of i) a front absorber made of carbon, concrete 
and steel of 10 interaction lengths 
($\lambda_{\rm I}$) located between the interaction point (IP) and the 
spectrometer that filters 
out hadrons, ii) a beam shield throughout its entire length, iii) 
a dipole magnet with a field integral of 3 T$\cdot$m,
iv) five tracking stations, each composed of 
two planes of cathode pad chambers, v) two trigger stations, each 
equipped with two planes of resistive plate chambers and vi) an iron wall of 
7.2~$\lambda_{\rm I}$ placed between the tracking and trigger systems. 
The following detectors are also involved in the analysis.
The Silicon Pixel Detector (SPD), which constitutes the two innermost layers 
of the Inner Tracking System (with pseudo-rapidity coverage 
$\vert \eta_{\rm lab} \vert < 2$ and $\vert \eta_{\rm lab} \vert <1.4$ 
for the inner and outer layer, respectively), is used for reconstructing 
the position of the collision point.
Two scintillator arrays (V0) placed on each side of the 
IP (with pseudo-rapidity coverage $2.8 <\eta_{\rm lab} < 5.1$ and 
$-3.7 <\eta_{\rm lab} <-1.7$) are used for triggering purposes and to  
reject offline beam-induced background events. 
The V0 as well as the two T0 arrays, made of quartz 
Cherenkov counters and covering the 
acceptance $4.6 < \eta_{\rm lab} < 4.9$ and $-3.3 < \eta_{\rm lab} < -3.0$,
are employed to determine the luminosity.
The Zero Degree Calorimeters (ZDC) located at 112.5~m on both sides of the IP 
are also used in the offline event selection.

The results presented in this Letter are based on the data samples recorded 
by ALICE during the 2013 p--Pb run. Due to the different energy 
per nucleon of the colliding beams 
($E_{\rm p}$ = 4 TeV, $E_{\rm Pb}$ = 1.58 TeV), the centre-of-mass of the 
nucleon-nucleon collisions is shifted in rapidity by $\Delta y$ = 0.465 with 
respect to the laboratory frame in the direction of the proton beam. Data 
were collected with two beam configurations by reversing the rotation 
direction of the p and Pb beams. This allowed us to measure muon production
in the rapidity intervals 
$2.03 < y_{\rm cms} < 3.53$ and $-4.46 < y_{\rm cms} < -2.96$, 
the positive rapidities corresponding to the 
proton beam traveling in the direction of the muon spectrometer 
(p--Pb configuration) and the negative rapidities to the opposite 
case (Pb--p configuration).

The data samples used for the analysis consist of muon-triggered events, 
requiring in addition to the minimum bias (MB) trigger condition the presence 
of one candidate track with a \pt~above a threshold value 
in the muon trigger system. 
The MB trigger is formed by a coincidence between signals in the two V0 
arrays ($> 99$\% efficiency for the selection of non-single diffractive 
collisions). Data were collected using two different 
trigger \pt~thresholds, of about 0.5 GeV/$c$ and 4.2 GeV/$c$, defined as the 
$p_{\rm T}$ value for which the muon trigger probability is 50\%.
In the following, the low- and high-\pt~trigger threshold samples are 
referred to as MSL and MSH, respectively.
The beam-induced background events were removed by using the 
timing information from the V0 arrays. Collisions outside the nominal 
timing of the LHC bunches were rejected using the information from the ZDC. 
The maximum instantaneous luminosity at the ALICE IP during data-taking was 
$10^{29}~{\rm Hz/cm}^2$, and the probability for multiple interactions 
in a bunch crossing (pile-up) was at most 2\%. 
The integrated luminosities for the used
data samples are 
$196 \pm 7\ \mu\rm b^{-1}\ 
(4.9 \cdot 10^3 \pm 0.2 \cdot 10^3 \mu\rm b^{-1})$ 
in the p--Pb configuration and $254 \pm 9\ \mu\rm b^{-1}\ 
(5.8 \cdot 10^3 \pm 0.2 \cdot 10^3 \mu\rm b^{-1})$ 
in the Pb--p configuration for MSL- (MSH-) triggered events. 
The calculation of the 
integrated luminosities and associated uncertainties is discussed in
Section~\ref{sec:strat}.

%% file: c03Analysis.tex
\section{Data analysis}\label{sec:strat}

\subsection{Muon candidate selection}\label{subsec:selec}
               
The offline selection criteria of muon candidates are similar to those 
described in~\cite{Abelev:2012pi,Abelev:2012qh}. Tracks were required to be 
reconstructed in the kinematic region $-4 < \eta_{\rm lab} < -2.5$ 
and $170^\circ < \theta_{\rm abs} < 178^\circ$ ($\theta_{\rm abs}$ is the 
polar angle at the end of the absorber). 
In addition, tracks in the tracking system were required to match track 
segments in the trigger system. This results in a very effective rejection of 
the hadronic background that is absorbed in the iron wall.
A selection on the Distance of Closest Approach (DCA) to the primary vertex of 
each track weighted with its momentum ($p$) was also applied. 
The maximum value is set to $6 \sigma_{p \cdot {\rm DCA}}$, 
where $\sigma_{p \cdot {\rm DCA}}$ is the resolution on this quantity.
This latter further reduces the contribution from fake tracks coming from the 
association of uncorrelated clusters in the tracking chambers 
and beam-induced background tracks. 
The measurement of muons from heavy-flavour hadron decays is performed in 
the interval $2 < p_{\rm T} < 16$~GeV/$c$ by combining MSL-triggered and 
MSH-triggered events.
The former are used up to~\pt~= 7~GeV/$c$, the latter at higher~\pt.
The large yield of muons from secondary light-hadron decays produced inside 
the front absorber prevents the measurement below $p_{\rm T} = 2$~GeV/$c$.
In the \pt~interval of the measurement, the background 
contribution consists mainly of muons from decays of primary charged pions 
and charged kaons produced at the interaction point.
The component of muons from J/$\psi$ decays, found to be less 
than 1--3\% of the inclusive muon yield, 
depending on rapidity and \pt, was not subtracted. 
Moreover, the background contribution of muons from W and Z/$\gamma^\star$ 
is also small in the \pt~interval of interest~\cite{Alice:2016wka} (less 
than 2--3\% at $p_{\rm T}$ = 16 GeV/$c$). 

\subsection{Analysis strategy}\label{subsec:strat}

Nuclear matter effects on the production of muons from heavy-flavour 
hadron decays can be quantified by means of the nuclear modification 
factor, 
$R_{\rm pPb}^{\mu^\pm \leftarrow {\rm {HF}}}$, which can be written as: 
\begin{linenomath}
\begin{equation}
R_{\rm pPb}^{\mu^\pm\leftarrow {\rm {HF}}} (p_{\rm T}) =
{1\over A} \cdot 
{{\rm d}\sigma_{\rm {pPb}}^{\mu^\pm\leftarrow {\rm {HF}}}/{\rm d}p_{\rm T}
\over
{{\rm d}\sigma_{\rm {pp}}^{\mu^\pm\leftarrow {\rm {HF}}}/
{\rm d}p_{\rm T}}},
\label{eq:RpPbDef}
\end{equation}
\end{linenomath}
where $A$ is the mass number of the Pb nucleus, 
${{\rm d}\sigma_{\rm {pp}}^{\mu^\pm\leftarrow {\rm {HF}}}/{\rm d}p_{\rm T}}$ 
and
${\rm d}\sigma_{\rm {pPb}}^{\mu^\pm\leftarrow {\rm {HF}}}/{\rm d}p_{\rm T}$ are
the \pt-differential production cross sections of muons from heavy-flavour 
hadron decays in pp and p--Pb collisions, respectively. 

The latter is evaluated as:
\begin{linenomath}
\begin{equation}
{{{\rm d}\sigma_{\rm pPb}^{\mu^\pm\leftarrow {\rm {HF}}}} \over 
{{\rm d}p_{\rm T}}} = 
\left({{{\rm d}N_{\rm pPb}^{\mu^\pm}} \over {{\rm d}p_{\rm T}}} - 
{{{\rm d}N_{\rm pPb}^{\mu^\pm\leftarrow {\rm \pi,K}}} \over 
{{\rm d}p_{\rm T}}}\right) \cdot 
{1\over {L_{\rm int}}},
\label{eq:CrossSecHF}
\end{equation}
\end{linenomath}
where ${\rm d}N^{\mu^\pm} / {\rm d}p_{\rm T}$ and 
${\rm d}N^{\mu^\pm\leftarrow {\rm \pi,K}} / {\rm d}p_{\rm T}$ are 
the \pt-differential yields of inclusive muons and of muons 
from charged-pion and charged-kaon decays, respectively. 
The integrated luminosity 
$L_{\rm int}$ is computed as $N_{\rm MB}$/$\sigma_{\rm MB}$, 
where $N_{\rm MB}$ and $\sigma_{\rm MB}$ are the number of MB 
collisions and the 
MB trigger cross section, respectively. 
The latter was measured in van der Meer scans 
and is 
$2.09 \pm 0.07$~b ($2.12 \pm 0.07$~b) for the p--Pb (Pb--p) 
configuration~\cite{Abelev:2014epa}. 
Since the analysis is based on muon-triggered events, the number of equivalent 
MB events is evaluated as 
$N_{\rm MB} = F_{\rm MSL (MSH)} \cdot N_{\rm MSL (MSH)}$, where 
$N_{\rm MSL (MSH)}$  is the number of analysed MSL- (MSH-) triggered events, 
and $F_{\rm MSL (MSH)}$ is a normalisation factor.
The number of MSL- and MSH- triggered events amounts to 
$1.45 \cdot 10^7 (2.63 \cdot 10^7)$ and $10^7 (1.53 \cdot 10^7)$ for
the p--Pb (Pb--p) samples, respectively.
The normalisation factor is 
determined with two different procedures described hereafter.
The first procedure is based on the 
offline selection of muon-triggered events in the MB data sample. 
In this approach, $F_{\rm MSL}$ is the inverse of the probability of 
meeting the MSL trigger condition in an MB event. 
The normalisation factor $F_{\rm MSH}$ is obtained as the inverse of the 
product of the probability of meeting the MSH trigger condition 
in a MSL event and that of meeting the MSL trigger condition 
in a MB event. 
The second procedure is based on the run-averaged ratio of the MB trigger rate 
to that of muon triggers (MSL or MSH), each corrected by the fraction of 
events passing the event-selection criteria.
Note that in both procedures, the number of MB events 
is corrected for pile-up.
Finally, the weighted average of the results 
obtained with the two approaches is computed, using the
statistical uncertainty as weight. 
The results are $F_{\rm MSL} = 28.20 \pm 0.08\ (20.50 \pm 0.04)$ and 
$F_{\rm MSH} = 1032.8 \pm 7.2\ (798.3 \pm 4.8)$ at 
forward (backward) rapidity. The quoted uncertainties are statistical.

The measured \pt-differential muon yield is corrected for acceptance and 
for the tracking and trigger efficiencies using the same procedure as for the 
analysis of pp collisions at $\sqrt {s}$ = 2.76 and 
7 TeV~\cite{Abelev:2012pi,Abelev:2012qh}. 
This procedure is based on a Monte Carlo simulation using as input the 
\pt~and rapidity distributions of muons from beauty-hadron decays predicted
by Fixed Order Next To Leading Log (FONLL) 
calculations~\cite{Cacciari:2012ny}\footnote{The sensitivity of the 
product of acceptance and efficiency on the 
input distributions was estimated 
by comparing the results with those from a simulation using muons from 
charm decays. The differences are negligible (less than 1\%).}. 
The detector description and its response are modelled using the GEANT3 
transport package~\cite{Brun:1994aa} taking into account the time evolution 
of the detector configuration. For $p_{\rm T} > 2$~GeV/$c$, 
the product of acceptance and efficiency in MSL-triggered events tends to 
saturate at a value close to 85\% and 75\% at forward (p--Pb configuration) 
and backward rapidity (Pb--p configuration), 
respectively. The lower value obtained for the Pb--p system is mainly due to a 
lower efficiency of the tracking chambers in the corresponding data 
taking period. 
The MSH trigger efficiency plateau is 
only just reached at \pt~= 16 GeV/$c$, which leads to values of 
the acceptance times efficiency slightly 
lower than those obtained for the MSL trigger, even in the high \pt~region.

The subtraction of background muons from charged-pion and charged-kaon decays 
is based on a data-tuned Monte Carlo cocktail.
First, the contribution of muons from charged-pion and charged-kaon decays in 
$2.03 < y_{\rm cms} < 3.53$ is estimated 
by extrapolating to forward rapidity the \pt-differential yields per 
minimum-bias event of charged pions and kaons measured by the 
ALICE Collaboration in the rapidity region $-0.5 < y_{\rm cms} < 0$ for 
$p_{\rm T}$ values up to $p_{\rm T} = 20$~GeV/$c$~\cite{Adam:2016dau}.
A further \pt~extrapolation, by means 
of a power-law fit, was performed to extend the \pt~coverage to 
the charged-pion and charged-kaon momentum range, which is relevant to 
estimate the contribution of muons from charged-pion 
and charged-kaon decays up to \pt~=~16 GeV/$c$.

The rapidity extrapolation of the 
$\lbrack{\rm d}^2 N^{\pi^\pm, K^\pm} / 
{\rm d}p_{\rm T} {\rm d} y\rbrack_{{\rm mid}- y_{\rm cms}}$ 
mid-rapidity charged-pion 
and charged-kaon yields to forward rapidity is performed according to:
\begin{linenomath}
\begin{equation}
\label{fextrap}
 \frac {{\rm d}^2 N^{\pi^\pm, K^\pm}} 
{{\rm d}p_{\rm T} {\rm d} y} = 
F_{\rm extrap} (p_{\rm T}, y) \cdot
 \bigg\lbrack \frac {
{\rm d}^2 N^{\pi^\pm, K^\pm}} {{\rm d}p_{\rm T} {\rm d} y}
\bigg\rbrack _{{\rm mid}-y_{\rm cms}}
\end{equation} 
\end{linenomath}
where the \pt- and y-dependent 
extrapolation factor $F_{\rm extrap}(p_{\rm T}, y)$ is obtained by 
means of the DPMJET event generator~\cite{Roesler:2000he}, which 
describes the pseudo-rapidity distribution of charged particles in 
$-2 < \eta_{\rm lab} < 2$ reasonably well~\cite{ALICE:2012xs}.
The HIJING 2.1 generator~\cite{Xu:2012au} 
is employed to estimate the 
systematic uncertainty (Section~\ref{sec:syst}). 
It was also checked that compatible 
results are obtained with the AMPT model~\cite{Lin:2004en}. 
Then, the (\pt, $y$) 
distributions of muons from charged-pion 
and charged-kaon decays in the acceptance of the muon spectrometer are 
generated with a simulation, using as input the charged-pion and charged-kaon 
distributions obtained with the extrapolation procedure described above.
The absorber effect is 
accounted for by rejecting charged pions and charged kaons that do not 
decay within 
a distance corresponding to one hadronic interaction length in the absorber. 
The charged-pion and charged-kaon 
distributions at backward rapidity, for $-4.46 < y_{\rm cms} < -2.96$, 
are estimated by using the distributions extrapolated at forward 
rapidity with DPMJET as a starting point, as discussed above.
These \pt~and $y$ distributions are 
scaled by the \pt-dependent charged-particle asymmetry factor 
measured by the CMS Collaboration for
$1.3 < \vert \eta_{\rm cms} \vert < 1.8$~\cite{Khachatryan:2015xaa}. 
The systematic uncertainty resulting from the different rapidity coverage
is discussed in Section~\ref{sec:syst}.
Finally, the distributions of muons from charged-pion and charged-kaon decays 
at backward rapidity are obtained with the fast simulation procedure described 
above for the forward rapidity 
region. 
The obtained yields per event of muons from charged-pion and charged-kaon 
decays at forward and backward rapidities 
are then scaled by $N_{\rm MB}$ and subtracted from the inclusive 
muon yields.

The relative contribution to the inclusive muon yield due to muons
from charged-pion and charged-kaon decays decreases 
with increasing \pt~from about 27\% (35\%) at \pt~= 2 GeV/$c$ to 2\% (2\%) at 
\pt~= 16 GeV/$c$, at 
forward (backward) rapidity. In the smaller overlapping acceptance 
$ 2.96 < \vert y_{\rm cms} \vert < 3.53$ used for 
the measurement of the forward-to-backward ratio 
$R_{\rm FB}^{\mu^\pm \leftarrow {\rm HF}}$, 
the background fraction decreases from about 19\% (41\%) at \pt~= 2 GeV/$c$ 
to 1\% (3\%) at \pt~= 16 GeV/$c$, at forward (backward) rapidity.

The \pt-differential cross sections of muons from heavy-flavour 
hadron decays in pp collisions at $\sqrt{s}$ = 5.02 TeV, needed 
for the computation of 
$R_{\rm pPb}$ at forward and backward rapidity, are obtained by applying a 
pQCD-driven energy and rapidity scaling 
to the measured 
\pt-differential cross sections in pp collisions at $\sqrt s$~=~7 TeV in the 
kinematic region $2.5 <y_{\rm cms} < 4.0$ and 
$2 <p_{\rm T}< 12$~GeV/$c$~\cite{Abelev:2012pi}. The 
scaling factor and its uncertainty are evaluated 
using FONLL calculations~\cite{Cacciari:2012ny}
with different sets of factorisation 
and renormalisation scales and quark masses, as detailed 
in~\cite{Averbeck:2011ga}. 
The current measurement of the pp \pt-differential cross section at 
$\sqrt s$~=~7~TeV is limited to $p_{\rm T} < 12$~GeV/$c$.
Therefore, the \pt-differential cross sections
in $12 <p_{\rm T} < 16$~GeV/$c$ at $\sqrt s$ = 5.02 TeV are obtained 
from FONLL calculations at $\sqrt s$ = 5.02 TeV, rescaled to 
match the result of the data-driven procedure in $6 < p_{\rm T} < 12$ GeV/$c$.
Note that in the limited 
interval $2 < p_{\rm T} < 10$~GeV/$c$, 
the \pt-differential cross section of muons from heavy-flavour hadron 
decays was also 
measured in pp collisions at $\sqrt s$ = 2.76 TeV. 
As a cross-check, it was verified that when using this measurement 
in the procedure for scaling to $\sqrt s$ = 5.02 TeV, 
compatible results are obtained with respect to 
those from the $\sqrt s$ = 7 TeV case, although with larger uncertainties
\footnote{This results from larger uncertainties and a larger energy gap at 
$\sqrt s$ = 2.76 TeV compared to $\sqrt s$ = 7 TeV.}.

The forward-to-backward ratio, 
$R_{\rm {FB}}^{\mu^\pm\leftarrow {\rm {HF}}}$, 
defined as the ratio of the cross section of muons from 
heavy-flavour hadron 
decays at forward rapidity to that at backward rapidity in a 
rapidity interval symmetric with respect to $y_{\rm cms} = 0$,
\begin{linenomath}
\begin{equation}
R_{\rm {FB}}^{\mu^\pm\leftarrow {\rm {HF}}} (p_{\rm T})= 
{\lbrack
{\rm d}\sigma_{\rm {pPb}}^{\mu^\pm\leftarrow {\rm {HF}}}/{\rm d}p_{\rm T}
\rbrack_{2.96 < y_{\rm cms} < 3.53} 
\over
\ \ \ \ \ \lbrack
{{\rm d}\sigma_{\rm {pPb}}^{\mu^\pm\leftarrow {\rm {HF}}}/{\rm d}p_{\rm T}}
\rbrack_{-3.53<  y_{\rm cms}  < -2.96}},
\label{eq:RFBDef}
\end{equation}
\end{linenomath}
is also a sensitive observable for the study of CNM effects.
This ratio can be computed only in the 
restricted overlapping $y$ interval 
$2.96 < \vert y_{\rm cms} \vert < 3.53$ covered at 
both forward and backward rapidity. 

\subsection{Systematic uncertainties}\label{sec:syst}

The measurement of the \pt-differential 
cross sections of muons from heavy-flavour hadron decays is affected by 
systematic uncertainties of the inclusive muon yield, the background 
subtraction and the determination of the integrated luminosity. For the 
nuclear modification factor, also the systematic uncertainty on the pp 
reference cross section must be considered.

The systematic uncertainty affecting the yield of inclusive muons contains 
the 2\% (3\%) systematic uncertainty on the muon tracking efficiency at 
forward (backward) rapidity~\cite{Abelev:2013yxa,Abelev:2014oea}
and the systematic uncertainty associated with the muon trigger 
efficiency of 1\% with the MSL trigger and 4\% with the MSH trigger. 
A detailed description of the procedure used to evaluate these 
uncertainties is found in~\cite{Abelev:2014ffa,Abelev:2013yxa,Abelev:2014oea}. 
A 0.5\% systematic uncertainty due to the efficiency of the matching 
between tracking and trigger information is also added. A conservative 
\pt-dependent systematic uncertainty of $0.5\% \cdot p_{\rm T}$ (in GeV/$c$) 
is assigned to take into account the difference between the  
true (unknown) residual mis-alignment of the spectrometer and the simulated 
one. 

The systematic uncertainty of the estimate of the yield of muons from 
charged-pion and charged-kaon decays contains contributions 
from the uncertainty on 
i) the measured mid-rapidity \pt~distributions of charged pions and 
kaons and their \pt~extrapolation, of 5--8\%, ii) the rapidity extrapolation, 
of 7--26\% (2--27\%) at 
forward (backward) rapidity, depending on \pt, estimated by comparing the 
results from DPMJET and HIJING generators and iii) the absorber 
effect, of 15\%, obtained by varying the interaction length in the absorber 
within reasonable limits. 
At backward rapidity, in addition to previous systematic uncertainties a 
systematic uncertainty arises from 
the procedure that makes use of the asymmetry factor measured by the CMS 
Collaboration~\cite{Khachatryan:2015xaa} 
in different rapidity intervals with respect to our 
measurement. This uncertainty, about 15--18\%, is calculated by varying the 
asymmetry factor between unity 
and two times the measured value for charged particles. An additional 15\% 
uncertainty is included to account for the variations with \pt~of the 
measured asymmetry factor with respect to a uniform distribution in the high 
\pt~region.
All the aforementioned uncertainties are added in quadrature to obtain 
the total uncertainty on the background subtraction, which results in an 
uncertainty on the \pt-differential cross section and nuclear modification 
factor of muons from heavy-flavour hadron decays of 1--7\% (1--15\%) at 
forward (backward) rapidity (Table~\ref{tsyst}).   

The systematic uncertainty of the measurement of the integrated 
luminosity includes contributions from $\sigma_{\rm MB}$ and $N_{\rm MB}$.
The systematic uncertainty of $N_{\rm MB}$ of about 1\% reflects the 
difference between the normalisation factor $F_{\rm MSL (MSH)}$ values 
obtained with the two different procedures described in 
Section~\ref{subsec:strat}.
The systematic uncertainty of $\sigma_{\rm MB}$ amounts to 3.5\% (3.2\%) for 
the p--Pb (Pb--p) configuration, with a total correlated uncertainty between 
these two configurations of 1.6\%. 
The luminosity measurement was performed independently by using a 
second reference cross section, based on particle detection by the T0 
detector~\cite{Abelev:2014epa}.  
The luminosities measured with the two detectors differ 
by at most 1\% throughout the whole data-taking period. This value is 
combined quadratically with the systematic uncertainties 
on $\sigma_{\rm MB}$ and $N_{\rm MB}$, leading to a total uncertainty on the 
integrated luminosity of 3.8\% 
(3.5\%) for the p--Pb (Pb--p) configuration.

The systematic uncertainty of the pp reference at $\sqrt s$ = 5.02 TeV 
accounts for the uncertainties of
i) the measurement of the \pt-differential cross 
section of muons from heavy-flavour hadron decays at $\sqrt s$ = 7 TeV, 
of 8--14\%, plus a global uncertainty of 3.5\% from the luminosity 
measurement~\cite{Abelev:2012sea} quoted separately,
ii) the energy scaling factor, obtained by considering different 
sets of factorisation and renormalisation scales and quark masses in FONLL 
as detailed in~\cite{Averbeck:2011ga}, of 3\% (7\%) at \pt~=~2 GeV/$c$ and 
2\% (4\%) at \pt~=~12 GeV/$c$ at forward (backward) rapidity, 
iii) the procedure based on FONLL predictions for $12 < p_{\rm T} < 
16$~GeV/$c$, of 
26\% (30\%) at forward (backward) rapidity, and iv) the rapidity extrapolation.
The uncertainty on the latter amounts to 2\% at forward rapidity and is 
negligible at backward rapidity.
It is estimated from the pp cross sections at $\sqrt s$ = 7 TeV measured 
in the full acceptance and in various rapidity 
sub-intervals~\cite{Abelev:2012pi}. These rapidity sub-intervals are combined 
in order to mimic the rapidity intervals investigated in the 
p--Pb and Pb--p configurations (Section~\ref{sec:det}),
scaled with FONLL to the full rapidity 
coverage and compared with the measurement. 

A summary of the systematic uncertainty sources previously 
discussed, after propagation to the measurements of
${\rm d}\sigma_{\rm pPb}^{\mu^\pm\leftarrow {\rm {HF}}}/{\rm d} p_{\rm T}$ and 
$R_{\rm pPb}^{\mu^\pm\leftarrow {\rm {HF}}}$, 
is presented in Table~\ref{tsyst}. 
The main contribution 
to the $R_{\rm pPb}^{\mu^\pm\leftarrow {\rm {HF}}}$ systematic uncertainty 
comes from the pp reference, in particular 
in the high \pt~region ($p_{\rm T} > 12$~GeV/$c$). 
Most of the systematic uncertainties are uncorrelated as a function of \pt, 
with the exception of the systematic uncertainties of mis-alignment in pp and 
p--Pb collisions which are correlated bin-to-bin in \pt, of the 
detector response which is partially correlated, 
and of the luminosity which is fully correlated. The total 
systematic uncertainty on $R_{\rm {pPb}}^{\mu^\pm\leftarrow {\rm {HF}}}$ 
varies within about 12--28\% (18--31\%) at forward (backward) rapidity. 

\begin{table}[!hbt] 
\centering
\begin{tabular}{c|c|c}
\hline
Source & Forward rapidity & Backward rapidity \\ 
\hline
Tracking efficiency & 2\% & 3\% \\
Trigger efficiency & 1\% (4\%) for MSL (MSH) & 1\% (4\%) for MSL (MSH) \\
Matching efficiency & 0.5\% & 0.5\% \\
Mis-alignment & 0.5$\%$$\cdot$\pt & 0.5$\%\cdot$\pt \\
Background subtraction & 1--7\% & 1--15\% \\
Integrated luminosity & 3.8\% & 3.5\% \\
\hline
$\sigma^{\mu^\pm \leftarrow {\rm HF}}_{\rm pp}$ ($p_{\rm T}$-dependent) & 9--26\% 
& 9--30\% \\
$\sigma^{\mu^\pm \leftarrow {\rm HF}}_{\rm pp}$ (global) & 3.5\% & 3.5\% \\
\hline
\end{tabular}
\caption{Systematic uncertainties affecting
the measurement of the \pt-differential cross section and nuclear modification 
factor of muons from heavy-flavour hadron decays at forward rapidity 
($2.03<y_{\rm cms}<3.53$) and backward rapidity 
($-4.46< y_{\rm cms}<-2.96$). See the text for details. 
For the \pt-dependent uncertainties, the minimum and maximum values are given. 
They are given at \pt~= 2 GeV/$c$ and \pt~= 16 GeV/$c$, 
except for the background subtraction where the first (last) value corresponds 
to \pt~= 16 (2) GeV/$c$. 
The systematic uncertainties of the pp reference 
($\sigma_{\rm {pp}}^{\rm \mu^\pm \leftarrow {\rm HF}}$ 
$p_{\rm T}$-dependent and global) contribute only to the 
systematic uncertainty on the nuclear modification factors.}
\label{tsyst}
\end{table}

All systematic uncertainties entering the 
${\rm d}\sigma_{{\rm pPb}}^{\mu^\pm\leftarrow {{\rm HF}}}/{\rm d}p_{\rm T}$ 
measurement at forward and backward rapidity affect the 
$R_{\rm {FB}}^{\mu^\pm\leftarrow {\rm {HF}}}$ measurement, with the exception 
of the 1.6\% contribution from the uncertainty on the luminosity, which is 
fully correlated between the results at forward and 
backward rapidity. 
The main contribution 
to the $R_{\rm {FB}}^{\mu^\pm\leftarrow {\rm {HF}}}$ systematic uncertainty 
comes from the muon background at low \pt~(\pt~$<$ 4~GeV/$c$) as well as the 
detector response and mis-alignment in the high-\pt~region. The total 
systematic uncertainty on $R_{\rm {FB}}^{\mu^\pm\leftarrow {\rm {HF}}}$ 
decreases with increasing \pt, 
from about 20\% (\pt~= 2 GeV/$c$) to 10\% (\pt~= 16 GeV/$c$).

%% file: c04Results.tex
\section{Results and comparison to model predictions}\label{sec:res}

\begin{figure}[!t]
\begin{center}
\includegraphics*[width=.9\textwidth]{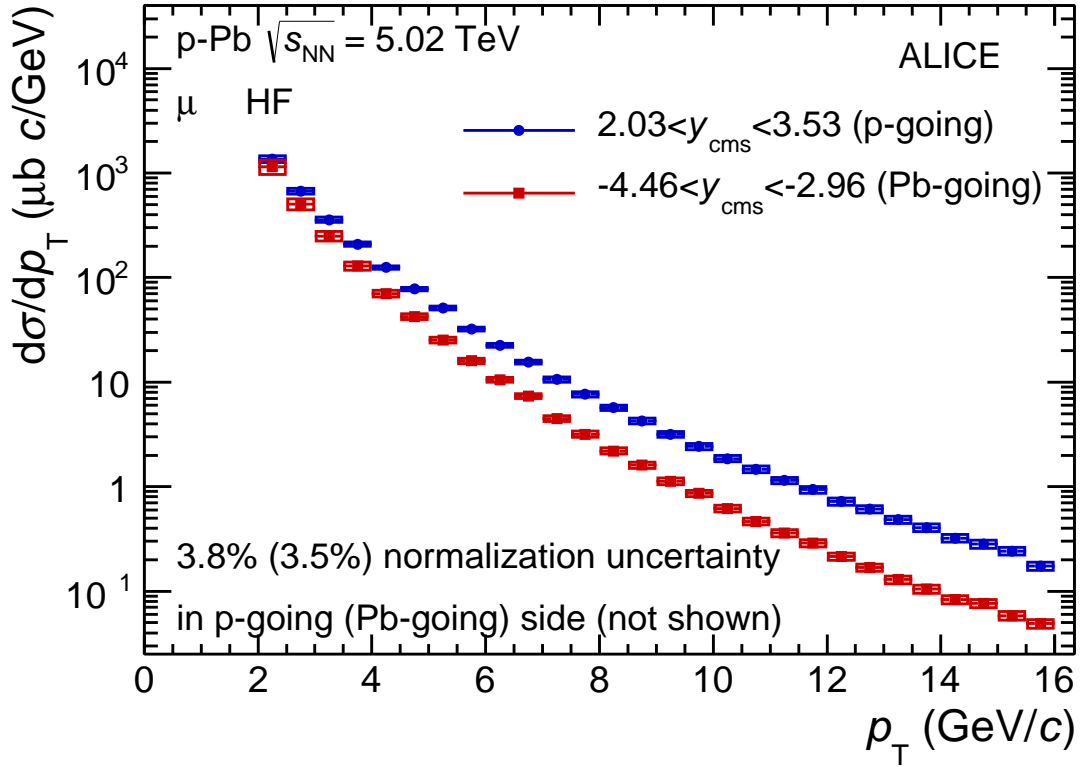}
\end{center}
\caption{Production cross sections of muons from heavy-flavour hadron 
decays as a function of \pt~for p--Pb collisions at 
$\sqrt {s_{\rm NN}}$~=~5.02~TeV 
at forward rapidity ($2.03<y_{\rm cms}<3.53$) and backward 
rapidity ($-4.46<y_{\rm cms}<-2.96$). Statistical uncertainties (bars) and 
systematic uncertainties (boxes) are shown. }
\label{fig:HFMxsec}
\end{figure}

\begin{figure}[!ht]
\begin{center}
\includegraphics*[width=0.85\textwidth]{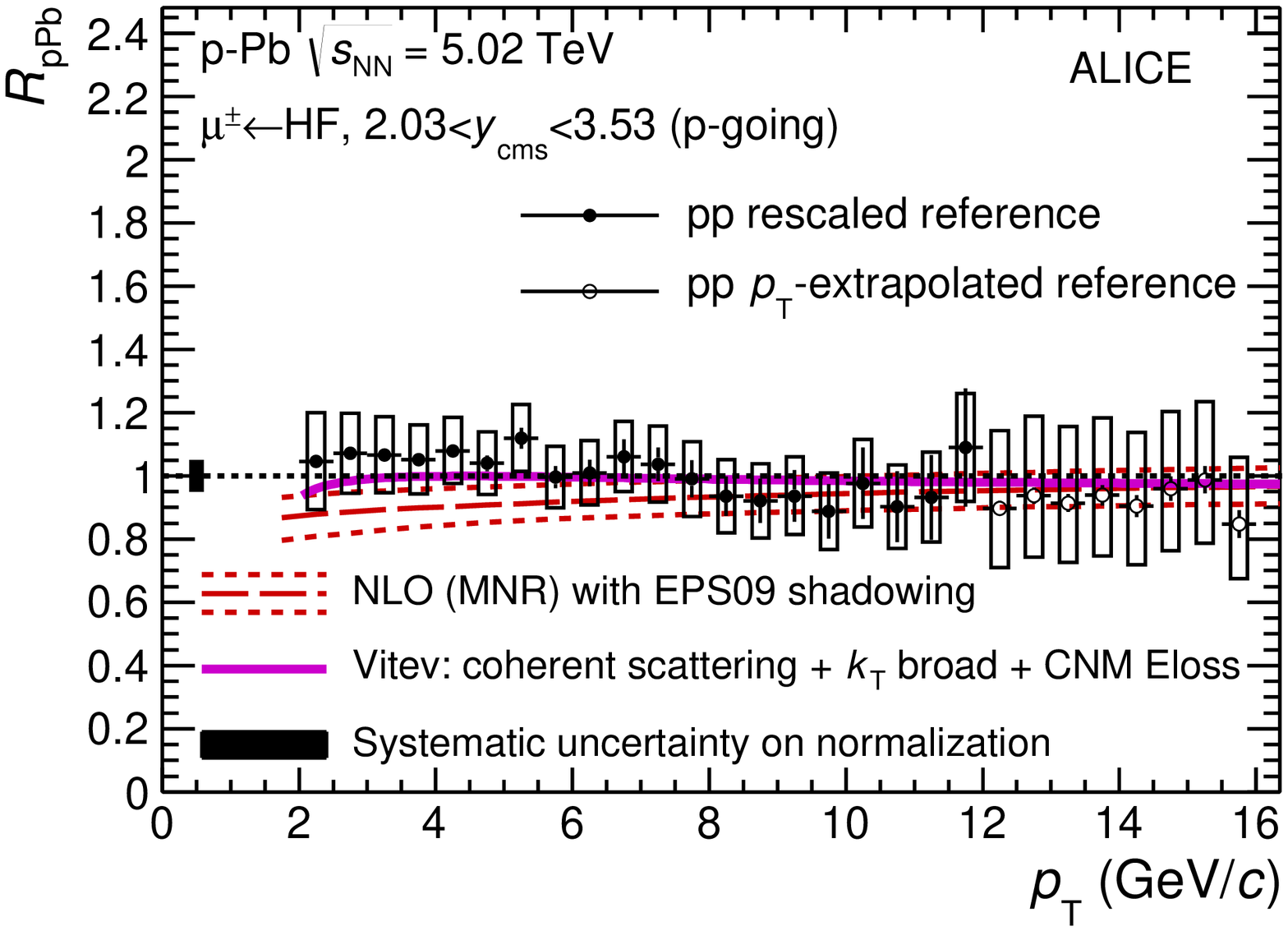}
\includegraphics*[width=0.85\textwidth]{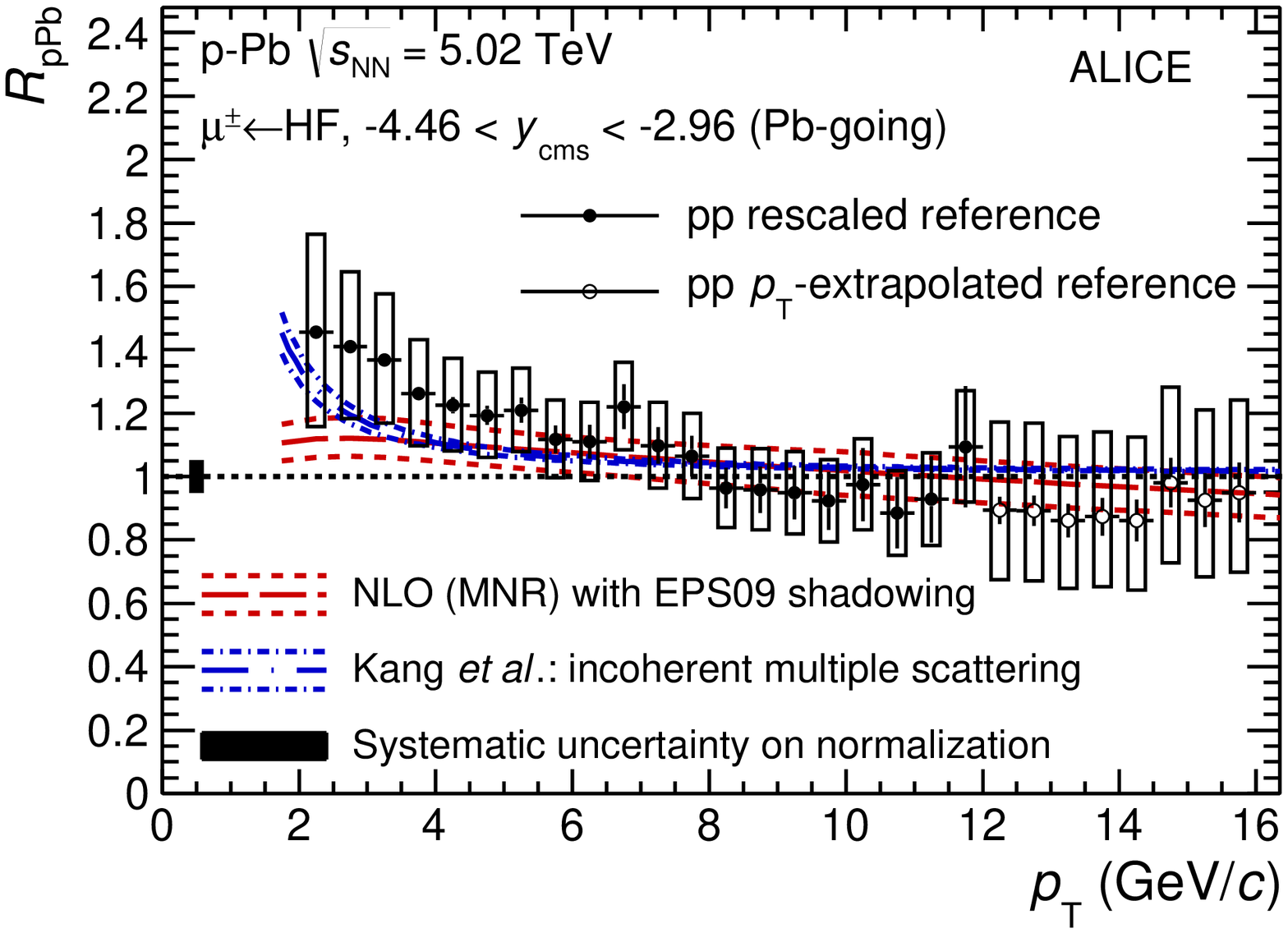}
\end{center}
\caption{Nuclear modification factor of muons from heavy-flavour hadron 
decays as a function of \pt~for p--Pb collisions at 
$\sqrt {s_{\rm NN}}$~=~5.02~TeV at forward rapidity ($2.03<y_{\rm cms}<3.53$, 
top) and backward rapidity ($-4.46<y_{\rm cms}<-2.96$, bottom) 
compared to model predictions~\cite{Mangano:1991jk, Sharma:2009hn, 
Kang:2014hha}. 
Statistical uncertainties (bars), 
systematic uncertainties (open boxes), and normalisation uncertainties 
(filled box at $R_{\rm pPb}^{\mu^\pm \leftarrow {\rm HF}}$ = 1) are shown.
Filled (open) symbols refer to the pp reference obtained from an 
energy and rapidity scaling to the measurement at $\sqrt s$ = 7 TeV 
(an extrapolation based on FONLL calculations).} 
\label{fig:ptRpPb}
\end{figure}

\begin{figure}[!ht]
\begin{center}
\includegraphics*[width=0.85\textwidth]{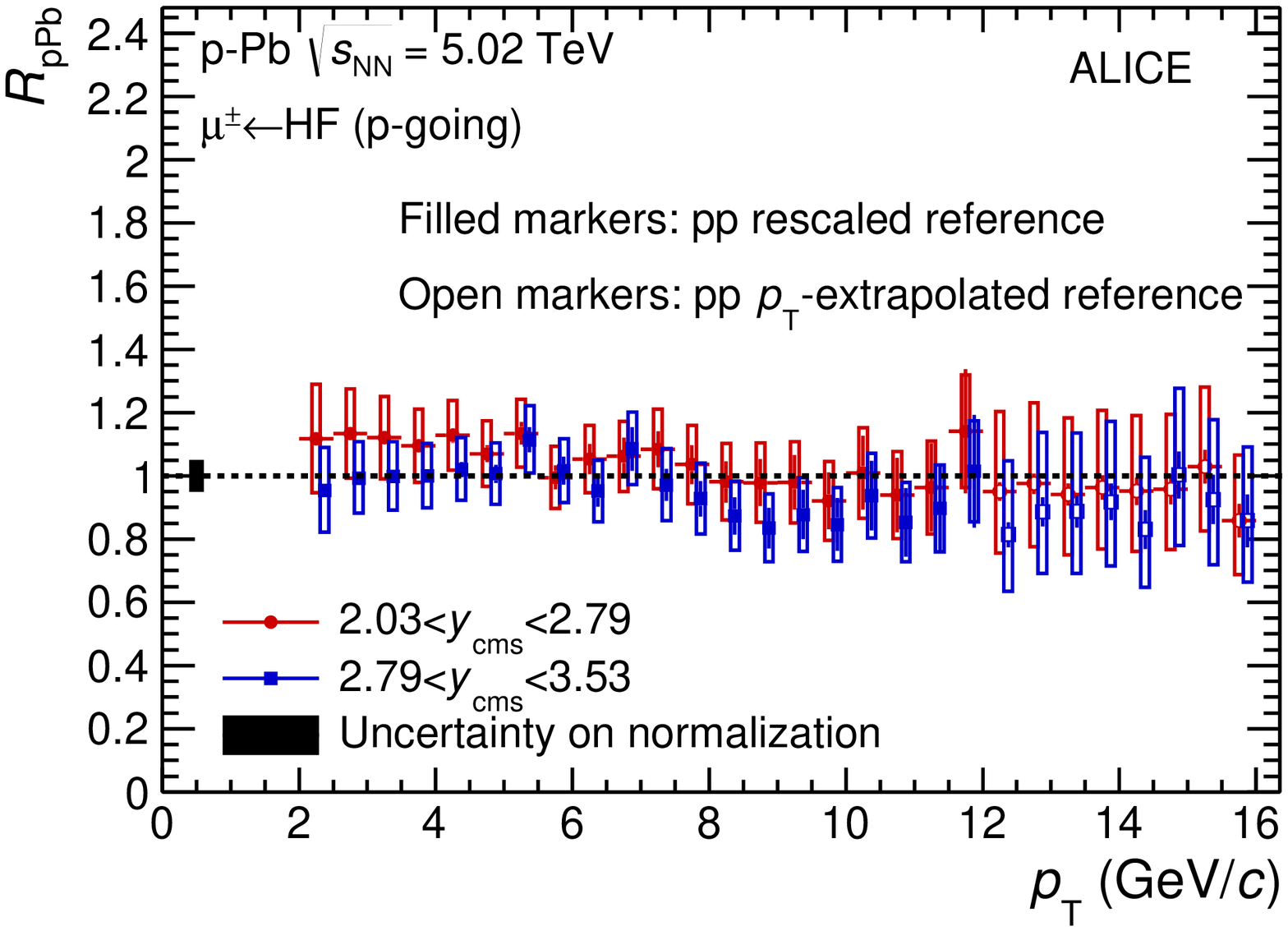}
\includegraphics*[width=0.85\textwidth]{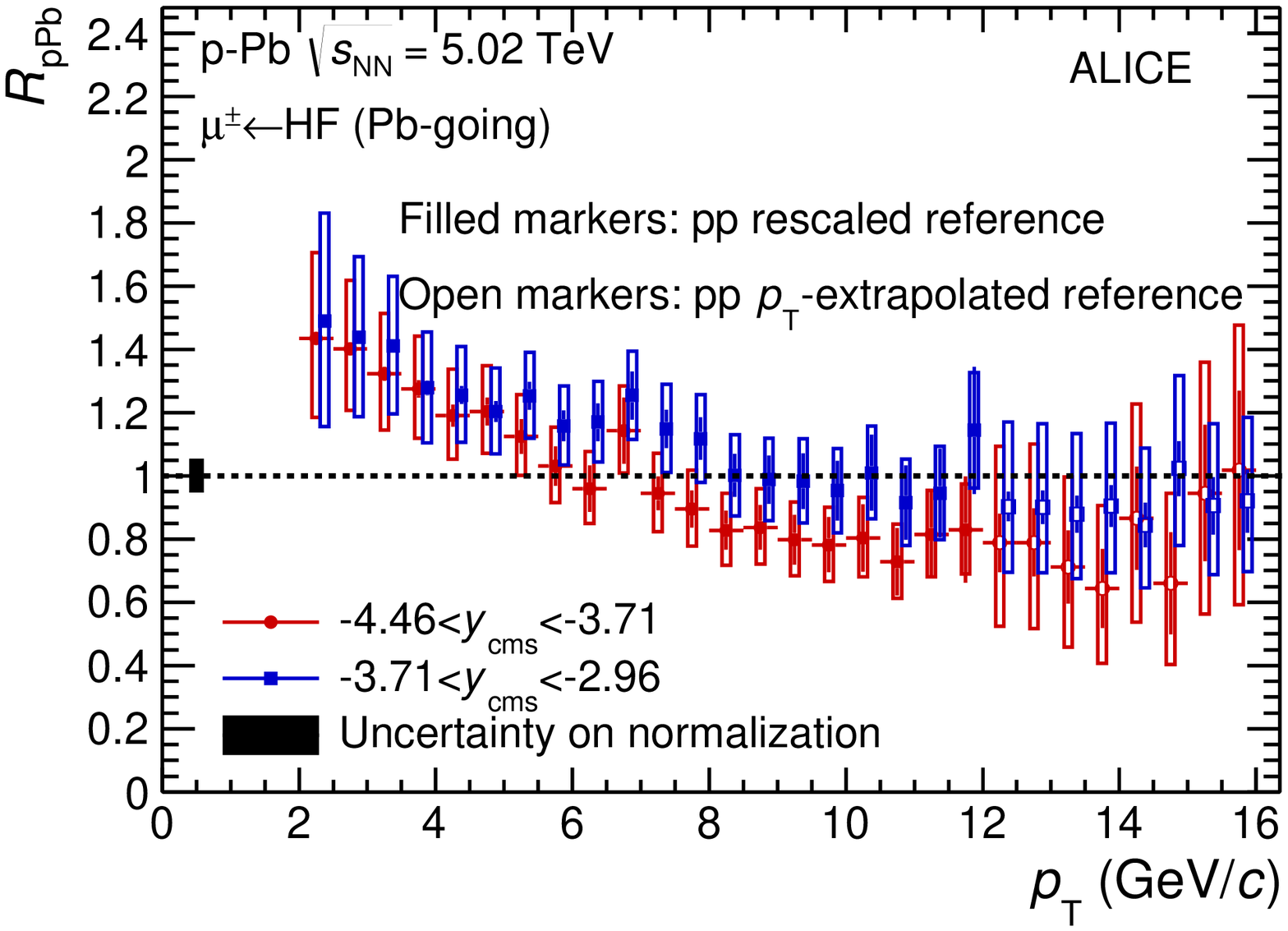}
\end{center}
\caption{Nuclear modification factors 
of muons from heavy-flavour hadron 
decays as a function of \pt~for p--Pb collisions at 
$\sqrt {s_{\rm NN}}$~=~5.02~TeV in two 
rapidity sub-intervals at forward (top) and 
backward (bottom) rapidity. 
Statistical uncertainties (bars), 
systematic uncertainties (open boxes), and normalisation uncertainties 
(filled box at $R_{\rm pPb}^{\mu^\pm \leftarrow {\rm HF}}$ = 1) are shown. 
For visibility, the points for the rapidity intervals 
$2.79 < y_{\rm cms} < 3.53$ and $-3.71 < y_{\rm cms} < -2.96$ are slightly 
shifted 
horizontally. Filled (open) symbols refer to the pp reference obtained from an 
energy and rapidity scaling to the measurement at $\sqrt s$ = 7 TeV 
(an extrapolation based on FONLL calculations).}
\label{fig:ycmsRpPb}
\end{figure}

The \pt-differential cross sections of muons from heavy-flavour hadron
decays measured in p--Pb collisions at $\sqrt {s_{\rm NN}}$ = 
5.02 TeV at forward rapidity ($2.03 <y_{\rm cms} < 3.53$) 
and backward rapidity ($-4.46 <y_{\rm cms} < 2.96$) in the interval 
$2 < p_{\rm T} < 16$~GeV/$c$
are displayed in Fig.~\ref{fig:HFMxsec}. They are further used to compute the 
nuclear modification factor $R_{\rm pPb}$. Vertical bars represent the 
statistical uncertainties and empty boxes, smaller than the symbols, 
the systematic uncertainties that 
include all sources discussed in Section~\ref{sec:strat}, 
except the normalisation 
uncertainties. These conventions related to the drawing of 
uncertainties apply also to the figures discussed in the following.

Figure~\ref{fig:ptRpPb} 
shows the \pt-differential nuclear modification factor, 
$R_{\rm pPb}^{\mu^\pm \leftarrow {\rm HF}}$, 
in p--Pb collisions at $\sqrt {s_{\rm NN}}$ = 5.02 TeV at 
forward rapidity (top panel) and backward rapidity 
(bottom panel). 
Besides statistical and systematic uncertainties, also the 
normalisation is shown as a filled box at 
$R_{\rm pPb}^{\mu^\pm \leftarrow {\rm HF}}$ = 1. The significantly 
smaller statistical (and larger systematic) 
uncertainties for $p_{\rm T} > 12$ GeV/$c$ 
compared to the interval $7< p_{\rm T} < 12$ GeV/$c$ 
reflect the different procedure used 
for the determination of the pp reference, described in 
Section~\ref{subsec:strat}. The \pt-differential 
$R_{\rm pPb}^{\mu^\pm \leftarrow {\rm HF}}$ at forward rapidity 
is compatible with unity within 
uncertainties over the whole \pt~range. 
At backward rapidity, $R_{\rm pPb}^{\mu^\pm \leftarrow {\rm HF}}$ is larger 
than unity with a maximum significance of $2.2\sigma$ for 
the interval $2.5 < p_{\rm T} < 3.5$~GeV/$c$, as calculated from the 
combined statistical 
and systematic uncertainties. 
At higher \pt, it is compatible with unity.
The measurements indicate that CNM 
effects are small and that the strong suppression of the yields of 
muons from heavy-flavour hadron decays observed in the 10\% most 
central Pb--Pb collisions~\cite{Abelev:2012qh} 
should result from final-state effects, e.g. the heavy-quark in-medium 
energy loss. 
The trends measured by ALICE in p--Pb collisions, including the hint for an 
enhancement at backward rapidity, are similar to those observed by 
the PHENIX Collaboration at RHIC for muons from heavy-flavour hadron decays 
measured in d--Au collisions at $\sqrt {s_{\rm NN}}$ = 200 GeV at forward 
($1.4 < y_{\rm cms} < 2.0$) and backward 
($-2.0 < y_{\rm cms} < -1.4$) rapidity~\cite{Adare:2013lkk}.

As shown in Fig.~\ref{fig:ptRpPb}, Next-to-Leading Order (NLO) perturbative 
QCD calculations by Mangano, Nason and Ridolfi (MNR)~\cite{Mangano:1991jk},
which make use of the EPS09~\cite{Eskola:2009uj} 
parameterization of nuclear PDFs 
(CTEQ6M~\cite{Stump:2003yu}) and do not include any final-state effect, 
describe the measurements in the two rapidity regions reasonably well
within experimental and theoretical uncertainties.
The data at forward 
rapidity are also well described by calculations including 
nuclear shadowing, $k_{\rm T}$ broadening and energy loss in cold nuclear 
matter~\cite{Sharma:2009hn}, which predict $R_{\rm pPb}$ very close to unity
over the whole momentum range of the measurement. 
An agreement with these calculations 
was also reported by ALICE for D mesons and 
electrons from heavy-flavour hadron decays measured at 
mid-rapidity~\cite{Abelev:2014hha,Adam:2015qda}.
The \pt-differential $R_{\rm pPb}^{\mu^\pm \leftarrow {\rm HF}}$ at backward 
rapidity is also compared with predictions from a model including incoherent 
multiple scattering effects of hard partons in the Pb nucleus 
both in initial-state and final-state 
interactions~\cite{Kang:2014hha}. 
This model expects also a 
small enhancement at low values of transverse momentum and describes the 
measurement fairly well over the whole \pt~range.
The same model is able to describe both
the \pt-differential $R_{\rm pPb}$ of 
electrons from heavy-flavour hadron decays measured at mid-rapidity with 
ALICE, which is also consistent with unity within 
uncertainties~\cite{Adam:2015qda},
and the enhancement 
seen at backward rapidity in d--Au collisions at 
$\sqrt{s_{\rm NN}}$ = 200 GeV for muons from heavy-flavour 
hadron decays~\cite{Kang:2014hha}. 
Theoretical calculations based on the Color Glass 
Condensate model~\cite{Fujii:2015lld} 
predict that for the rapidity interval $2.5 < y_{\rm cms} < 3.53$, the 
$R_{\rm pPb}$ of 
muons from charm-hadron decays for the interval $0 < p_{\rm T} < 4$~GeV/$c$ 
increases with increasing \pt~from about 0.6 to 0.85. 
This predicted $R_{\rm pPb}$ is slightly 
smaller than that reported 
here for muons from heavy-flavour hadron decays\footnote{For the 
interval $0 < p_{\rm T} < 4$~GeV/$c$ the component of muons from charm-hadron 
decays dominates according to FONLL calculations~\cite{Cacciari:2012ny}.}, 
although for a slightly different rapidity interval.

The \pt-differential nuclear modification factors of muons from heavy-flavour 
hadron decays were also studied as a function of rapidity, 
by dividing each of the two intervals in two sub-intervals. 
The results are presented in Fig.~\ref{fig:ycmsRpPb}. In both the 
forward (top panel) and backward (bottom panel) rapidity regions, no 
significant difference is observed 
between the nuclear modification factors measured in the two rapidity 
sub-intervals \footnote{It cannot be excluded that a degree of 
correlation between the two rapidity sub-intervals, difficult 
to quantify, is present in the various systematic uncertainty sources.}.

\begin{figure}[!t]
\begin{center}
\includegraphics*[width=0.9\textwidth]{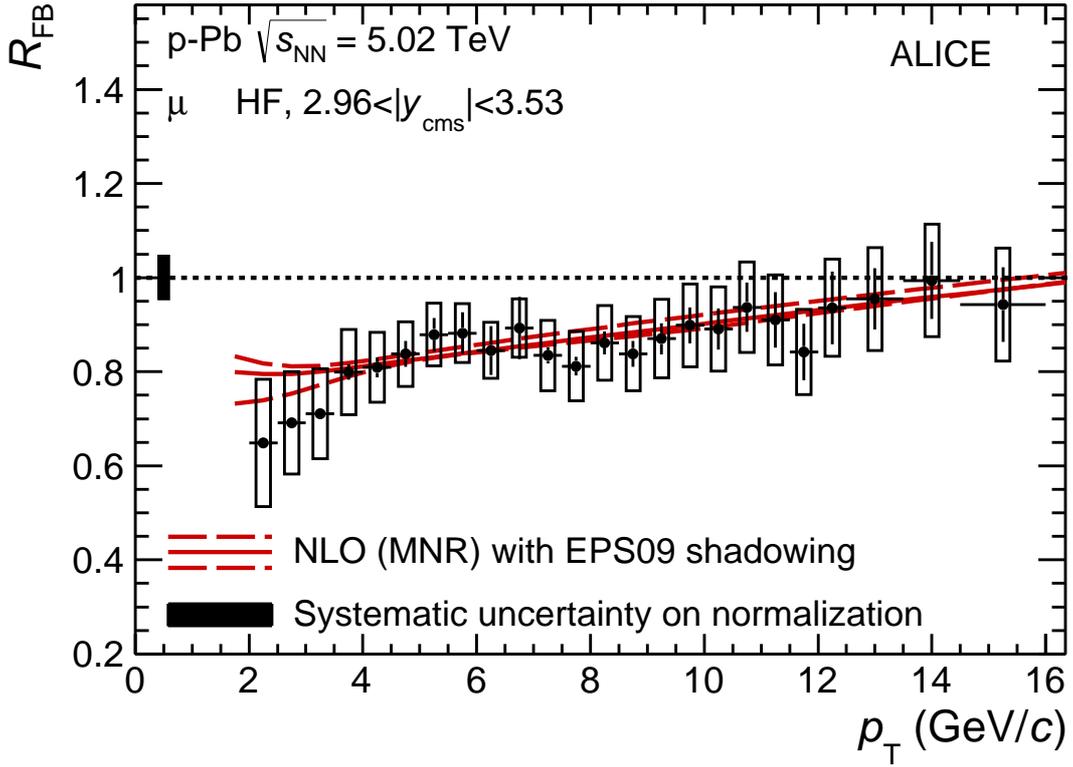}
\end{center}
\caption{Forward-to-backward ratio of muons from heavy-flavour hadron 
decays as a fucntion of \pt~for p--Pb collisions at 
$\sqrt {s_{\rm NN}}$~=~5.02~TeV compared to 
model predictions~\cite{Mangano:1991jk}. 
Statistical uncertainties (bars), 
systematic uncertainties (open boxes) and normalisation uncertainties 
(filled box at $R_{\rm FB}^{\mu^\pm \leftarrow {\rm HF}}$ = 1) are shown.} 
\label{fig:ptRFB}
\end{figure}
Figure~\ref{fig:ptRFB} shows $R_{\rm FB}^{\mu^\pm \leftarrow {\rm HF}}$ for 
muons from heavy-flavour hadron decays for the rapidity region 
$2.96 < \vert y_{\rm cms} \vert < 3.53$ function of \pt~(Eq.~\ref{eq:RFBDef}). 
The forward-to-backward ratio is found to be smaller 
than unity at intermediate \pt, 
with a significance of $3.7\sigma$ for $2.5 < p_{\rm T} < 3.5$~GeV/$c$, 
and it rises gradually towards unity 
with increasing \pt. This observable is also well described by NLO pQCD 
calculations with the EPS09 modification of the CTEQ6M PDFs.

%% file: c05Conclusion.tex
\section{Conclusion}\label{sec:concl}

In summary, the production of muons from heavy-flavour hadron decays has been 
measured in p--Pb collisions at $\sqrt {s_{\rm NN}}$ = 5.02 TeV for 
$2 < p_{\rm T} < 16$~GeV/c with the ALICE 
detector at the CERN LHC. Measurements of the production 
cross sections and nuclear 
modification factors have been presented as a function of $p_{\rm T}$ 
at forward 
($2.03 <y_{\rm cms} < 3.53$, p-going direction) and backward 
($-4.46 <y_{\rm cms} < -2.96$, Pb-going direction) rapidity. 
Moreover, the $p_{\rm T}$-differential forward-to-backward ratio has been 
also studied in the smaller 
overlapping interval $2.96 < \vert y_{\rm cms}\vert  <3.53$. 
At forward rapidity, the nuclear modification factor is 
compatible with unity over the whole \pt~range. 
At backward rapidity, 
a deviation from binary scaling is suggested in the interval 
$2.5 <p_{\rm T} < 3.5$~GeV/$c$ with a significance of about $2\sigma$.
The observed trends in the 
$R_{\rm pPb}^{\mu \leftarrow {\rm HF}}$ measurements are reflected in the 
forward-to-backward ratio, which shows a clear tendency to be below unity, 
with a deviation of $3.7\sigma$ for 
$2.5 <p_{\rm T} < 3.5$~GeV/$c$. 
The measured nuclear modification factors and the 
forward-to-backward ratio
are reproduced within uncertainties by NLO pQCD calculations including 
nuclear modification of the PDFs.
The nuclear modification factor at forward rapidity is in 
agreement with a model calculation including CNM effects based 
on a nuclear shadowing scenario, $k_{\rm T}$ 
broadening and energy loss in cold nuclear matter. 
The data at backward rapidity are also reproduced 
by a model including incoherent multiple scattering effects.
The results indicate that the suppression of the production of
high-\pt~muons from heavy-flavour hadron decays in the 0--10\% most 
central Pb--Pb collisions measured by ALICE is due to 
final-state effects induced by the hot and dense medium formed in 
these collisions.

%% file: fa_2017-01-25.tex

The ALICE Collaboration would like to thank all its engineers and technicians for their invaluable contributions to the construction of the experiment and the CERN accelerator teams for the outstanding performance of the LHC complex.
The ALICE Collaboration gratefully acknowledges the resources and support provided by all Grid centres and the Worldwide LHC Computing Grid (WLCG) collaboration.
The ALICE Collaboration acknowledges the following funding agencies for their support in building and running the ALICE detector:
A. I. Alikhanyan National Science Laboratory (Yerevan Physics Institute) Foundation (ANSL), State Committee of Science and World Federation of Scientists (WFS), Armenia;
Austrian Academy of Sciences and Nationalstiftung f\"{u}r Forschung, Technologie und Entwicklung, Austria;
Ministry of Communications and High Technologies, National Nuclear Research Center, Azerbaijan;
Conselho Nacional de Desenvolvimento Cient\'{\i}fico e Tecnol\'{o}gico (CNPq), Universidade Federal do Rio Grande do Sul (UFRGS), Financiadora de Estudos e Projetos (Finep) and Funda\c{c}\~{a}o de Amparo \`{a} Pesquisa do Estado de S\~{a}o Paulo (FAPESP), Brazil;
Ministry of Science \& Technology of China (MSTC), National Natural Science Foundation of China (NSFC) and Ministry of Education of China (MOEC) , China;
Ministry of Science, Education and Sport and Croatian Science Foundation, Croatia;
Ministry of Education, Youth and Sports of the Czech Republic, Czech Republic;
The Danish Council for Independent Research | Natural Sciences, the Carlsberg Foundation and Danish National Research Foundation (DNRF), Denmark;
Helsinki Institute of Physics (HIP), Finland;
Commissariat \`{a} l'Energie Atomique (CEA) and Institut National de Physique Nucl\'{e}aire et de Physique des Particules (IN2P3) and Centre National de la Recherche Scientifique (CNRS), France;
Bundesministerium f\"{u}r Bildung, Wissenschaft, Forschung und Technologie (BMBF) and GSI Helmholtzzentrum f\"{u}r Schwerionenforschung GmbH, Germany;
Ministry of Education, Research and Religious Affairs, Greece;
National Research, Development and Innovation Office, Hungary;
Department of Atomic Energy Government of India (DAE) and Council of Scientific and Industrial Research (CSIR), New Delhi, India;
Indonesian Institute of Science, Indonesia;
Centro Fermi - Museo Storico della Fisica e Centro Studi e Ricerche Enrico Fermi and Istituto Nazionale di Fisica Nucleare (INFN), Italy;
Institute for Innovative Science and Technology , Nagasaki Institute of Applied Science (IIST), Japan Society for the Promotion of Science (JSPS) KAKENHI and Japanese Ministry of Education, Culture, Sports, Science and Technology (MEXT), Japan;
Consejo Nacional de Ciencia (CONACYT) y Tecnolog\'{i}a, through Fondo de Cooperaci\'{o}n Internacional en Ciencia y Tecnolog\'{i}a (FONCICYT) and Direcci\'{o}n General de Asuntos del Personal Academico (DGAPA), Mexico;
Nationaal instituut voor subatomaire fysica (Nikhef), Netherlands;
The Research Council of Norway, Norway;
Commission on Science and Technology for Sustainable Development in the South (COMSATS), Pakistan;
Pontificia Universidad Cat\'{o}lica del Per\'{u}, Peru;
Ministry of Science and Higher Education and National Science Centre, Poland;
Korea Institute of Science and Technology Information and National Research Foundation of Korea (NRF), Republic of Korea;
Ministry of Education and Scientific Research, Institute of Atomic Physics and Romanian National Agency for Science, Technology and Innovation, Romania;
Joint Institute for Nuclear Research (JINR), Ministry of Education and Science of the Russian Federation and National Research Centre Kurchatov Institute, Russia;
Ministry of Education, Science, Research and Sport of the Slovak Republic, Slovakia;
National Research Foundation of South Africa, South Africa;
Centro de Aplicaciones Tecnol\'{o}gicas y Desarrollo Nuclear (CEADEN), Cubaenerg\'{\i}a, Cuba, Ministerio de Ciencia e Innovacion and Centro de Investigaciones Energ\'{e}ticas, Medioambientales y Tecnol\'{o}gicas (CIEMAT), Spain;
Swedish Research Council (VR) and Knut \& Alice Wallenberg Foundation (KAW), Sweden;
European Organization for Nuclear Research, Switzerland;
National Science and Technology Development Agency (NSDTA), Suranaree University of Technology (SUT) and Office of the Higher Education Commission under NRU project of Thailand, Thailand;
Turkish Atomic Energy Agency (TAEK), Turkey;
National Academy of  Sciences of Ukraine, Ukraine;
Science and Technology Facilities Council (STFC), United Kingdom;
National Science Foundation of the United States of America (NSF) and United States Department of Energy, Office of Nuclear Physics (DOE NP), United States of America.

%% file: Alice_Authorlist_2017-Jan-25_mod.tex


\begingroup
\small
\begin{flushleft}
S.~Acharya$^\textrm{\scriptsize 139}$,
D.~Adamov\'{a}$^\textrm{\scriptsize 87}$,
M.M.~Aggarwal$^\textrm{\scriptsize 91}$,
G.~Aglieri Rinella$^\textrm{\scriptsize 34}$,
M.~Agnello$^\textrm{\scriptsize 30}$,
N.~Agrawal$^\textrm{\scriptsize 47}$,
Z.~Ahammed$^\textrm{\scriptsize 139}$,
N.~Ahmad$^\textrm{\scriptsize 17}$,
S.U.~Ahn$^\textrm{\scriptsize 69}$,
S.~Aiola$^\textrm{\scriptsize 143}$
% ...
\renewcommand\labelenumi{\textsuperscript{\theenumi}~}

\section*{Affiliation notes}
\renewcommand\theenumi{\roman{enumi}}
\begin{Authlist}
\item \Adef{0}Deceased
\item \Adef{idp1752848}{Also at: Dipartimento DET del Politecnico di Torino, Turin, Italy}
\item \Adef{idp1772240}{Also at: Georgia State University, Atlanta, Georgia, United States}
\item \Adef{idp3971792}{Also at: M.V. Lomonosov Moscow State University, D.V. Skobeltsyn Institute of Nuclear, Physics, Moscow, Russia}
\item \Adef{idp4308080}{Also at: Department of Applied Physics, Aligarh Muslim University, Aligarh, India}
\end{Authlist}

\section*{Collaboration Institutes}
\renewcommand\theenumi{\arabic{enumi}~}

$^{1}$A.I. Alikhanyan National Science Laboratory (Yerevan Physics Institute) Foundation, Yerevan, Armenia
\\
$^{2}$Benem\'{e}rita Universidad Aut\'{o}noma de Puebla, Puebla, Mexico
\\
$^{3}$Bogolyubov Institute for Theoretical Physics, Kiev, Ukraine
\\
$^{4}$Bose Institute, Department of Physics 
and Centre for Astroparticle Physics and Space Science (CAPSS), Kolkata, India
\\
$^{5}$Budker Institute for Nuclear Physics, Novosibirsk, Russia
\\
$^{6}$California Polytechnic State University, San Luis Obispo, California, United States
\\
$^{7}$Central China Normal University, Wuhan, China
\\
$^{8}$Centre de Calcul de l'IN2P3, Villeurbanne, Lyon, France
\\
$^{9}$Centro de Aplicaciones Tecnol\'{o}gicas y Desarrollo Nuclear (CEADEN), Havana, Cuba
\\
$^{10}$Centro de Investigaciones Energ\'{e}ticas Medioambientales y Tecnol\'{o}gicas (CIEMAT), Madrid, Spain
\\
$^{11}$Centro de Investigaci\'{o}n y de Estudios Avanzados (CINVESTAV), Mexico City and M\'{e}rida, Mexico
\\
$^{12}$Centro Fermi - Museo Storico della Fisica e Centro Studi e Ricerche ``Enrico Fermi', Rome, Italy
\\
$^{13}$Chicago State University, Chicago, Illinois, United States
\\
$^{14}$China Institute of Atomic Energy, Beijing, China
\\
$^{15}$COMSATS Institute of Information Technology (CIIT), Islamabad, Pakistan
\\
$^{16}$Departamento de F\'{\i}sica de Part\'{\i}culas and IGFAE, Universidad de Santiago de Compostela, Santiago de Compostela, Spain
\\
$^{17}$Department of Physics, Aligarh Muslim University, Aligarh, India
\\
$^{18}$Department of Physics, Ohio State University, Columbus, Ohio, United States
\\
$^{19}$Department of Physics, Sejong University, Seoul, South Korea
\\
$^{20}$Department of Physics, University of Oslo, Oslo, Norway
\\
$^{21}$Department of Physics and Technology, University of Bergen, Bergen, Norway
\\
$^{22}$Dipartimento di Fisica dell'Universit\`{a} 'La Sapienza'
and Sezione INFN, Rome, Italy
\\
$^{23}$Dipartimento di Fisica dell'Universit\`{a}
and Sezione INFN, Cagliari, Italy
\\
$^{24}$Dipartimento di Fisica dell'Universit\`{a}
and Sezione INFN, Trieste, Italy
\\
$^{25}$Dipartimento di Fisica dell'Universit\`{a}
and Sezione INFN, Turin, Italy
\\
$^{26}$Dipartimento di Fisica e Astronomia dell'Universit\`{a}
and Sezione INFN, Bologna, Italy
\\
$^{27}$Dipartimento di Fisica e Astronomia dell'Universit\`{a}
and Sezione INFN, Catania, Italy
\\
$^{28}$Dipartimento di Fisica e Astronomia dell'Universit\`{a}
and Sezione INFN, Padova, Italy
\\
$^{29}$Dipartimento di Fisica `E.R.~Caianiello' dell'Universit\`{a}
and Gruppo Collegato INFN, Salerno, Italy
\\
$^{30}$Dipartimento DISAT del Politecnico and Sezione INFN, Turin, Italy
\\
$^{31}$Dipartimento di Scienze e Innovazione Tecnologica dell'Universit\`{a} del Piemonte Orientale and INFN Sezione di Torino, Alessandria, Italy
\\
$^{32}$Dipartimento Interateneo di Fisica `M.~Merlin'
and Sezione INFN, Bari, Italy
\\
$^{33}$Division of Experimental High Energy Physics, University of Lund, Lund, Sweden
\\
$^{34}$European Organization for Nuclear Research (CERN), Geneva, Switzerland
\\
$^{35}$Excellence Cluster Universe, Technische Universit\"{a}t M\"{u}nchen, Munich, Germany
\\
$^{36}$Faculty of Engineering, Bergen University College, Bergen, Norway
\\
$^{37}$Faculty of Mathematics, Physics and Informatics, Comenius University, Bratislava, Slovakia
\\
$^{38}$Faculty of Nuclear Sciences and Physical Engineering, Czech Technical University in Prague, Prague, Czech Republic
\\
$^{39}$Faculty of Science, P.J.~\v{S}af\'{a}rik University, Ko\v{s}ice, Slovakia
\\
$^{40}$Faculty of Technology, Buskerud and Vestfold University College, Tonsberg, Norway
\\
$^{41}$Frankfurt Institute for Advanced Studies, Johann Wolfgang Goethe-Universit\"{a}t Frankfurt, Frankfurt, Germany
\\
$^{42}$Gangneung-Wonju National University, Gangneung, South Korea
\\
$^{43}$Gauhati University, Department of Physics, Guwahati, India
\\
$^{44}$Helmholtz-Institut f\"{u}r Strahlen- und Kernphysik, Rheinische Friedrich-Wilhelms-Universit\"{a}t Bonn, Bonn, Germany
\\
$^{45}$Helsinki Institute of Physics (HIP), Helsinki, Finland
\\
$^{46}$Hiroshima University, Hiroshima, Japan
\\
$^{47}$Indian Institute of Technology Bombay (IIT), Mumbai, India
\\
$^{48}$Indian Institute of Technology Indore, Indore, India
\\
$^{49}$Indonesian Institute of Sciences, Jakarta, Indonesia
\\
$^{50}$Inha University, Incheon, South Korea
\\
$^{51}$Institut de Physique Nucl\'eaire d'Orsay (IPNO), Universit\'e Paris-Sud, CNRS-IN2P3, Orsay, France
\\
$^{52}$Institute for Nuclear Research, Academy of Sciences, Moscow, Russia
\\
$^{53}$Institute for Subatomic Physics of Utrecht University, Utrecht, Netherlands
\\
$^{54}$Institute for Theoretical and Experimental Physics, Moscow, Russia
\\
$^{55}$Institute of Experimental Physics, Slovak Academy of Sciences, Ko\v{s}ice, Slovakia
\\
$^{56}$Institute of Physics, Academy of Sciences of the Czech Republic, Prague, Czech Republic
\\
$^{57}$Institute of Physics, Bhubaneswar, India
\\
$^{58}$Institute of Space Science (ISS), Bucharest, Romania
\\
$^{59}$Institut f\"{u}r Informatik, Johann Wolfgang Goethe-Universit\"{a}t Frankfurt, Frankfurt, Germany
\\
$^{60}$Institut f\"{u}r Kernphysik, Johann Wolfgang Goethe-Universit\"{a}t Frankfurt, Frankfurt, Germany
\\
$^{61}$Institut f\"{u}r Kernphysik, Westf\"{a}lische Wilhelms-Universit\"{a}t M\"{u}nster, M\"{u}nster, Germany
\\
$^{62}$Instituto de Ciencias Nucleares, Universidad Nacional Aut\'{o}noma de M\'{e}xico, Mexico City, Mexico
\\
$^{63}$Instituto de F\'{i}sica, Universidade Federal do Rio Grande do Sul (UFRGS), Porto Alegre, Brazil
\\
$^{64}$Instituto de F\'{\i}sica, Universidad Nacional Aut\'{o}noma de M\'{e}xico, Mexico City, Mexico
\\
$^{65}$IRFU, CEA, Universit\'{e} Paris-Saclay, F-91191 Gif-sur-Yvette, France, Saclay, France
\\
$^{66}$iThemba LABS, National Research Foundation, Somerset West, South Africa
\\
$^{67}$Joint Institute for Nuclear Research (JINR), Dubna, Russia
\\
$^{68}$Konkuk University, Seoul, South Korea
\\
$^{69}$Korea Institute of Science and Technology Information, Daejeon, South Korea
\\
$^{70}$KTO Karatay University, Konya, Turkey
\\
$^{71}$Laboratoire de Physique Corpusculaire (LPC), Clermont Universit\'{e}, Universit\'{e} Blaise Pascal, CNRS--IN2P3, Clermont-Ferrand, France
\\
$^{72}$Laboratoire de Physique Subatomique et de Cosmologie, Universit\'{e} Grenoble-Alpes, CNRS-IN2P3, Grenoble, France
\\
$^{73}$Laboratori Nazionali di Frascati, INFN, Frascati, Italy
\\
$^{74}$Laboratori Nazionali di Legnaro, INFN, Legnaro, Italy
\\
$^{75}$Lawrence Berkeley National Laboratory, Berkeley, California, United States
\\
$^{76}$Moscow Engineering Physics Institute, Moscow, Russia
\\
$^{77}$Nagasaki Institute of Applied Science, Nagasaki, Japan
\\
$^{78}$National and Kapodistrian University of Athens, Physics Department, Athens, Greece, Athens, Greece
\\
$^{79}$National Centre for Nuclear Studies, Warsaw, Poland
\\
$^{80}$National Institute for Physics and Nuclear Engineering, Bucharest, Romania
\\
$^{81}$National Institute of Science Education and Research, Bhubaneswar, India
\\
$^{82}$National Nuclear Research Center, Baku, Azerbaijan
\\
$^{83}$National Research Centre Kurchatov Institute, Moscow, Russia
\\
$^{84}$Niels Bohr Institute, University of Copenhagen, Copenhagen, Denmark
\\
$^{85}$Nikhef, Nationaal instituut voor subatomaire fysica, Amsterdam, Netherlands
\\
$^{86}$Nuclear Physics Group, STFC Daresbury Laboratory, Daresbury, United Kingdom
\\
$^{87}$Nuclear Physics Institute, Academy of Sciences of the Czech Republic, \v{R}e\v{z} u Prahy, Czech Republic
\\
$^{88}$Oak Ridge National Laboratory, Oak Ridge, Tennessee, United States
\\
$^{89}$Petersburg Nuclear Physics Institute, Gatchina, Russia
\\
$^{90}$Physics Department, Creighton University, Omaha, Nebraska, United States
\\
$^{91}$Physics Department, Panjab University, Chandigarh, India
\\
$^{92}$Physics Department, University of Cape Town, Cape Town, South Africa
\\
$^{93}$Physics Department, University of Jammu, Jammu, India
\\
$^{94}$Physics Department, University of Rajasthan, Jaipur, India
\\
$^{95}$Physikalisches Institut, Eberhard Karls Universit\"{a}t T\"{u}bingen, T\"{u}bingen, Germany
\\
$^{96}$Physikalisches Institut, Ruprecht-Karls-Universit\"{a}t Heidelberg, Heidelberg, Germany
\\
$^{97}$Physik Department, Technische Universit\"{a}t M\"{u}nchen, Munich, Germany
\\
$^{98}$Purdue University, West Lafayette, Indiana, United States
\\
$^{99}$Pusan National University, Pusan, South Korea
\\
$^{100}$Research Division and ExtreMe Matter Institute EMMI, GSI Helmholtzzentrum f\"ur Schwerionenforschung GmbH, Darmstadt, Germany
\\
$^{101}$Rudjer Bo\v{s}kovi\'{c} Institute, Zagreb, Croatia
\\
$^{102}$Russian Federal Nuclear Center (VNIIEF), Sarov, Russia
\\
$^{103}$Saha Institute of Nuclear Physics, Kolkata, India
\\
$^{104}$School of Physics and Astronomy, University of Birmingham, Birmingham, United Kingdom
\\
$^{105}$Secci\'{o}n F\'{\i}sica, Departamento de Ciencias, Pontificia Universidad Cat\'{o}lica del Per\'{u}, Lima, Peru
\\
$^{106}$Sezione INFN, Bari, Italy
\\
$^{107}$Sezione INFN, Bologna, Italy
\\
$^{108}$Sezione INFN, Cagliari, Italy
\\
$^{109}$Sezione INFN, Catania, Italy
\\
$^{110}$Sezione INFN, Padova, Italy
\\
$^{111}$Sezione INFN, Rome, Italy
\\
$^{112}$Sezione INFN, Trieste, Italy
\\
$^{113}$Sezione INFN, Turin, Italy
\\
$^{114}$SSC IHEP of NRC Kurchatov institute, Protvino, Russia
\\
$^{115}$Stefan Meyer Institut f\"{u}r Subatomare Physik (SMI), Vienna, Austria
\\
$^{116}$SUBATECH, IMT Atlantique, Universit\'{e} de Nantes, CNRS-IN2P3, Nantes, France
\\
$^{117}$Suranaree University of Technology, Nakhon Ratchasima, Thailand
\\
$^{118}$Technical University of Ko\v{s}ice, Ko\v{s}ice, Slovakia
\\
$^{119}$Technical University of Split FESB, Split, Croatia
\\
$^{120}$The Henryk Niewodniczanski Institute of Nuclear Physics, Polish Academy of Sciences, Cracow, Poland
\\
$^{121}$The University of Texas at Austin, Physics Department, Austin, Texas, United States
\\
$^{122}$Universidad Aut\'{o}noma de Sinaloa, Culiac\'{a}n, Mexico
\\
$^{123}$Universidade de S\~{a}o Paulo (USP), S\~{a}o Paulo, Brazil
\\
$^{124}$Universidade Estadual de Campinas (UNICAMP), Campinas, Brazil
\\
$^{125}$Universidade Federal do ABC, Santo Andre, Brazil
\\
$^{126}$University of Houston, Houston, Texas, United States
\\
$^{127}$University of Jyv\"{a}skyl\"{a}, Jyv\"{a}skyl\"{a}, Finland
\\
$^{128}$University of Liverpool, Liverpool, United Kingdom
\\
$^{129}$University of Tennessee, Knoxville, Tennessee, United States
\\
$^{130}$University of the Witwatersrand, Johannesburg, South Africa
\\
$^{131}$University of Tokyo, Tokyo, Japan
\\
$^{132}$University of Tsukuba, Tsukuba, Japan
\\
$^{133}$University of Zagreb, Zagreb, Croatia
\\
$^{134}$Universit\'{e} de Lyon, Universit\'{e} Lyon 1, CNRS/IN2P3, IPN-Lyon, Villeurbanne, Lyon, France
\\
$^{135}$Universit\'{e} de Strasbourg, CNRS, IPHC UMR 7178, F-67000 Strasbourg, France, Strasbourg, France
\\
$^{136}$Universit\`{a} degli Studi di Pavia, Pavia, Italy
\\
$^{137}$Universit\`{a} di Brescia, Brescia, Italy
\\
$^{138}$V.~Fock Institute for Physics, St. Petersburg State University, St. Petersburg, Russia
\\
$^{139}$Variable Energy Cyclotron Centre, Kolkata, India
\\
$^{140}$Warsaw University of Technology, Warsaw, Poland
\\
$^{141}$Wayne State University, Detroit, Michigan, United States
\\
$^{142}$Wigner Research Centre for Physics, Hungarian Academy of Sciences, Budapest, Hungary
\\
$^{143}$Yale University, New Haven, Connecticut, United States
\\
$^{144}$Yonsei University, Seoul, South Korea
\\
$^{145}$Zentrum f\"{u}r Technologietransfer und Telekommunikation (ZTT), Fachhochschule Worms, Worms, Germany
\endgroup

%% file: AliRpPbHFM.bbl
\providecommand{\href}[2]{#2}\begingroup\raggedright\begin{thebibliography}{10}

\bibitem{Karsch:2006xs}
F.~Karsch, ``{Lattice simulations of the thermodynamics of strongly interacting
  elementary particles and the exploration of new phases of matter in
  relativistic heavy ion collisions},''
  \href{http://dx.doi.org/10.1088/1742-6596/46/1/017}{{\em J. Phys. Conf. Ser.}
  {\bfseries 46} (2006) 122--131},
\href{http://arxiv.org/abs/hep-lat/0608003}{{\ttfamily arXiv:hep-lat/0608003
  [hep-lat]}}.

\bibitem{Bazavov:2011nk}
A.~Bazavov {\em et~al.}, ``{The chiral and deconfinement aspects of the QCD
  transition},'' \href{http://dx.doi.org/10.1103/PhysRevD.85.054503}{{\em Phys.
  Rev.} {\bfseries D85} (2012) 054503},
\href{http://arxiv.org/abs/1111.1710}{{\ttfamily arXiv:1111.1710 [hep-lat]}}.

\bibitem{Miller:2007ri}
M.~L. Miller, K.~Reygers, S.~J. Sanders, and P.~Steinberg, ``{Glauber modeling
  in high energy nuclear collisions},''
  \href{http://dx.doi.org/10.1146/annurev.nucl.57.090506.123020}{{\em Ann. Rev.
  Nucl. Part. Sci.} {\bfseries 57} (2007) 205--243},
\href{http://arxiv.org/abs/nucl-ex/0701025}{{\ttfamily arXiv:nucl-ex/0701025
  [nucl-ex]}}.

\bibitem{Abelev:2012pi}
{\bfseries ALICE} Collaboration, B.~Abelev {\em et~al.}, ``{Heavy flavour decay
  muon production at forward rapidity in proton--proton collisions at $\sqrt{s}
  = 7$ TeV},'' \href{http://dx.doi.org/10.1016/j.physletb.2012.01.063}{{\em
  Phys. Lett.} {\bfseries B708} (2012) 265--275},
\href{http://arxiv.org/abs/1201.3791}{{\ttfamily arXiv:1201.3791 [hep-ex]}}.

\bibitem{Abelev:2012qh}
{\bfseries ALICE} Collaboration, B.~Abelev {\em et~al.}, ``{Production of muons
  from heavy flavour decays at forward rapidity in pp and Pb--Pb collisions at
  $\sqrt {s_{\rm NN}}$ = 2.76 TeV},''
  \href{http://dx.doi.org/10.1103/PhysRevLett.109.112301}{{\em Phys. Rev.
  Lett.} {\bfseries 109} (2012) 112301},
\href{http://arxiv.org/abs/1205.6443}{{\ttfamily arXiv:1205.6443 [hep-ex]}}.

\bibitem{Abelev:2012xe}
{\bfseries ALICE} Collaboration, B.~Abelev {\em et~al.}, ``{Measurement of
  electrons from semileptonic heavy-flavour hadron decays in pp collisions at
  $\sqrt{s}$ = 7 TeV},''
  \href{http://dx.doi.org/10.1103/PhysRevD.86.112007}{{\em Phys. Rev.}
  {\bfseries D86} (2012) 112007},
\href{http://arxiv.org/abs/1205.5423}{{\ttfamily arXiv:1205.5423 [hep-ex]}}.

\bibitem{Abelev:2012tca}
{\bfseries ALICE} Collaboration, B.~Abelev {\em et~al.}, ``{$\rm {D_{s}}^+$
  meson production at central rapidity in proton--proton collisions at
  $\sqrt{s}=7$ TeV},''
  \href{http://dx.doi.org/10.1016/j.physletb.2012.10.049}{{\em Phys. Lett.}
  {\bfseries B718} (2012) 279--294},
\href{http://arxiv.org/abs/1208.1948}{{\ttfamily arXiv:1208.1948 [hep-ex]}}.

\bibitem{Abelev:2012vra}
{\bfseries ALICE} Collaboration, B.~Abelev {\em et~al.}, ``{Measurement of
  charm production at central rapidity in proton-proton collisions at
  $\sqrt{s}=2.76$ TeV},'' \href{http://dx.doi.org/10.1007/JHEP07(2012)191}{{\em
  JHEP} {\bfseries 07} (2012) 191},
\href{http://arxiv.org/abs/1205.4007}{{\ttfamily arXiv:1205.4007 [hep-ex]}}.

\bibitem{Abelev:2012sca}
{\bfseries ALICE} Collaboration, B.~Abelev {\em et~al.}, ``{Measurement of
  electrons from beauty hadron decays in pp collisions at $\sqrt{s}=7$ TeV},''
  \href{http://dx.doi.org/10.1016/j.physletb.2013.01.069}{{\em Phys. Lett.}
  {\bfseries B721} (2013) 13--23},
\href{http://arxiv.org/abs/1208.1902}{{\ttfamily arXiv:1208.1902 [hep-ex]}}.

\bibitem{Abelev:2014hla}
{\bfseries ALICE} Collaboration, B.~Abelev {\em et~al.}, ``{Beauty production
  in pp collisions at $\sqrt{s}$ = 2.76 TeV measured via semi-electronic
  decays},'' \href{http://dx.doi.org/10.1016/j.physletb.2014.09.026}{{\em Phys.
  Lett.} {\bfseries B738} (2014) 97--108},
\href{http://arxiv.org/abs/1405.4144}{{\ttfamily arXiv:1405.4144 [nucl-ex]}}.

\bibitem{Abelev:2014gla}
{\bfseries ALICE} Collaboration, B.~Abelev {\em et~al.}, ``{Measurement of
  electrons from semileptonic heavy-flavor hadron decays in pp collisions at
  $\sqrt{s} = 2.76$ TeV},''
  \href{http://dx.doi.org/10.1103/PhysRevD.91.012001}{{\em Phys. Rev.}
  {\bfseries D91} no.~1, (2015) 012001},
\href{http://arxiv.org/abs/1405.4117}{{\ttfamily arXiv:1405.4117 [nucl-ex]}}.

\bibitem{ALICE:2012ab}
{\bfseries ALICE} Collaboration, B.~Abelev {\em et~al.}, ``{Suppression of high
  transverse momentum D mesons in central Pb--Pb collisions at $\sqrt{s_{\rm
  NN}}=2.76$ TeV},'' \href{http://dx.doi.org/10.1007/JHEP09(2012)112}{{\em
  JHEP} {\bfseries 09} (2012) 112},
\href{http://arxiv.org/abs/1203.2160}{{\ttfamily arXiv:1203.2160 [nucl-ex]}}.

\bibitem{Adam:2015sza}
{\bfseries ALICE} Collaboration, J.~Adam {\em et~al.}, ``{Transverse momentum
  dependence of D-meson production in Pb--Pb collisions at $
  \sqrt{{\mathrm{s}}_{\mathrm{NN}}}=$ 2.76 TeV},''
  \href{http://dx.doi.org/10.1007/JHEP03(2016)081}{{\em JHEP} {\bfseries 03}
  (2016) 081},
\href{http://arxiv.org/abs/1509.06888}{{\ttfamily arXiv:1509.06888 [nucl-ex]}}.

\bibitem{Adam:2016khe}
{\bfseries ALICE} Collaboration, J.~Adam {\em et~al.}, ``{Measurement of the
  production of high-$p_{\rm T}$ electrons from heavy-flavour hadron decays in
  Pb--Pb collisions at $\mathbf{\sqrt{\it s_{\rm{NN}}}}$ = 2.76 TeV},''
\href{http://arxiv.org/abs/1609.07104}{{\ttfamily arXiv:1609.07104 [nucl-ex]}}.

\bibitem{Abelev:2013lca}
{\bfseries ALICE} Collaboration, B.~Abelev {\em et~al.}, ``{D meson elliptic
  flow in non-central Pb--Pb collisions at $\sqrt{s_{\rm NN}}$ = 2.76TeV},''
  \href{http://dx.doi.org/10.1103/PhysRevLett.111.102301}{{\em Phys. Rev.
  Lett.} {\bfseries 111} (2013) 102301},
\href{http://arxiv.org/abs/1305.2707}{{\ttfamily arXiv:1305.2707 [nucl-ex]}}.

\bibitem{Abelev:2014ipa}
{\bfseries ALICE} Collaboration, B.~B. Abelev {\em et~al.}, ``{Azimuthal
  anisotropy of D meson production in Pb--Pb collisions at $\sqrt{s_{\rm NN}} =
  2.76$ TeV},'' \href{http://dx.doi.org/10.1103/PhysRevC.90.034904}{{\em Phys.
  Rev.} {\bfseries C90} no.~3, (2014) 034904},
\href{http://arxiv.org/abs/1405.2001}{{\ttfamily arXiv:1405.2001 [nucl-ex]}}.

\bibitem{Adam:2016ssk}
{\bfseries ALICE} Collaboration, J.~Adam {\em et~al.}, ``{Elliptic flow of
  electrons from heavy-flavour hadron decays at mid-rapidity in Pb--Pb
  collisions at $ \sqrt{{\mathrm{s}}_{\mathrm{NN}}}=2.76 $ TeV},''
  \href{http://dx.doi.org/10.1007/JHEP09(2016)028}{{\em JHEP} {\bfseries 09}
  (2016) 028},
\href{http://arxiv.org/abs/1606.00321}{{\ttfamily arXiv:1606.00321 [nucl-ex]}}.

\bibitem{Adam:2015pga}
{\bfseries ALICE} Collaboration, J.~Adam {\em et~al.}, ``{Elliptic flow of
  muons from heavy-flavour hadron decays at forward rapidity in Pb--Pb
  collisions at $\sqrt{s_{\rm NN}}= 2.76$ TeV},''
  \href{http://dx.doi.org/10.1016/j.physletb.2015.11.059}{{\em Phys. Lett.}
  {\bfseries B753} (2016) 41--56},
\href{http://arxiv.org/abs/1507.03134}{{\ttfamily arXiv:1507.03134 [nucl-ex]}}.

\bibitem{He:2014cla}
M.~He, R.~J. Fries, and R.~Rapp, ``{Heavy Flavor at the Large Hadron Collider
  in a Strong Coupling Approach},''
  \href{http://dx.doi.org/10.1016/j.physletb.2014.05.050}{{\em Phys. Lett.}
  {\bfseries B735} (2014) 445--450},
\href{http://arxiv.org/abs/1401.3817}{{\ttfamily arXiv:1401.3817 [nucl-th]}}.

\bibitem{Nahrgang:2013xaa}
M.~Nahrgang, J.~Aichelin, P.~B. Gossiaux, and K.~Werner, ``{Influence of
  hadronic bound states above $T\_c$ on heavy-quark observables in Pb + Pb
  collisions at at the CERN Large Hadron Collider},''
  \href{http://dx.doi.org/10.1103/PhysRevC.89.014905}{{\em Phys. Rev.}
  {\bfseries C89} no.~1, (2014) 014905},
\href{http://arxiv.org/abs/1305.6544}{{\ttfamily arXiv:1305.6544 [hep-ph]}}.

\bibitem{Uphoff:2012gb}
J.~Uphoff, O.~Fochler, Z.~Xu, and C.~Greiner, ``{Open Heavy Flavor in Pb+Pb
  Collisions at $\sqrt{s}=2.76$ TeV within a Transport Model},''
  \href{http://dx.doi.org/10.1016/j.physletb.2012.09.069}{{\em Phys. Lett.}
  {\bfseries B717} (2012) 430--435},
\href{http://arxiv.org/abs/1205.4945}{{\ttfamily arXiv:1205.4945 [hep-ph]}}.

\bibitem{Wicks:2005gt}
S.~Wicks, W.~Horowitz, M.~Djordjevic, and M.~Gyulassy, ``{Elastic, inelastic,
  and path length fluctuations in jet tomography},''
  \href{http://dx.doi.org/10.1016/j.nuclphysa.2006.12.048}{{\em Nucl. Phys.}
  {\bfseries A784} (2007) 426--442},
\href{http://arxiv.org/abs/nucl-th/0512076}{{\ttfamily arXiv:nucl-th/0512076
  [nucl-th]}}.

\bibitem{Horowitz:2011wm}
W.~A. Horowitz, ``{Testing pQCD and AdS/CFT Energy Loss at RHIC and LHC},''
  \href{http://dx.doi.org/10.1063/1.3700710}{{\em AIP Conf. Proc.} {\bfseries
  1441} (2012) 889--891},
\href{http://arxiv.org/abs/1108.5876}{{\ttfamily arXiv:1108.5876 [hep-ph]}}.

\bibitem{Lang:2012cx}
T.~Lang, H.~van Hees, J.~Steinheimer, G.~Inghirami, and M.~Bleicher, ``{Heavy
  quark transport in heavy ion collisions at energies available at the BNL
  Relativistic Heavy Ion Collider and at the CERN Large Hadron Collider within
  the UrQMD hybrid model},''
  \href{http://dx.doi.org/10.1103/PhysRevC.93.014901}{{\em Phys. Rev.}
  {\bfseries C93} no.~1, (2016) 014901},
\href{http://arxiv.org/abs/1211.6912}{{\ttfamily arXiv:1211.6912 [hep-ph]}}.

\bibitem{Cao:2013ita}
S.~Cao, G.-Y. Qin, and S.~A. Bass, ``{Heavy-quark dynamics and hadronization in
  ultrarelativistic heavy-ion collisions: Collisional versus radiative energy
  loss},'' \href{http://dx.doi.org/10.1103/PhysRevC.88.044907}{{\em Phys. Rev.}
  {\bfseries C88} no.~4, (2013) 044907},
\href{http://arxiv.org/abs/1308.0617}{{\ttfamily arXiv:1308.0617 [nucl-th]}}.

\bibitem{Eskola:2009uj}
K.~J. Eskola, H.~Paukkunen, and C.~A. Salgado, ``{EPS09: A New Generation of
  NLO and LO Nuclear Parton Distribution Functions},''
  \href{http://dx.doi.org/10.1088/1126-6708/2009/04/065}{{\em JHEP} {\bfseries
  04} (2009) 065},
\href{http://arxiv.org/abs/0902.4154}{{\ttfamily arXiv:0902.4154 [hep-ph]}}.

\bibitem{Helenius:2012wd}
I.~Helenius, K.~J. Eskola, H.~Honkanen, and C.~A. Salgado, ``{Impact-Parameter
  Dependent Nuclear Parton Distribution Functions: EPS09s and EKS98s and Their
  Applications in Nuclear Hard Processes},''
  \href{http://dx.doi.org/10.1007/JHEP07(2012)073}{{\em JHEP} {\bfseries 07}
  (2012) 073},
\href{http://arxiv.org/abs/1205.5359}{{\ttfamily arXiv:1205.5359 [hep-ph]}}.

\bibitem{Fujii:2013yja}
H.~Fujii and K.~Watanabe, ``{Heavy quark pair production in high energy pA
  collisions: Open heavy flavors},''
  \href{http://dx.doi.org/10.1016/j.nuclphysa.2013.10.006}{{\em Nucl. Phys.}
  {\bfseries A920} (2013) 78--93},
\href{http://arxiv.org/abs/1308.1258}{{\ttfamily arXiv:1308.1258 [hep-ph]}}.

\bibitem{Albacete:2012xq}
J.~L. Albacete, A.~Dumitru, H.~Fujii, and Y.~Nara, ``{CGC predictions for p+Pb
  collisions at the LHC},''
  \href{http://dx.doi.org/10.1016/j.nuclphysa.2012.09.012}{{\em Nucl. Phys.}
  {\bfseries A897} (2013) 1--27},
\href{http://arxiv.org/abs/1209.2001}{{\ttfamily arXiv:1209.2001 [hep-ph]}}.

\bibitem{Lev:1983hh}
M.~Lev and B.~Petersson, ``{Nuclear Effects at Large Transverse Momentum in a
  {QCD} Parton Model},''
\href{http://dx.doi.org/10.1007/BF01648792}{{\em Z. Phys.} {\bfseries C21}
  (1983) 155}.

\bibitem{Wang:1998ww}
X.-N. Wang, ``{Systematic study of high $p_{\rm T}$ hadron spectra in p p, pA
  and AA collisions from SPS to RHIC energies},''
  \href{http://dx.doi.org/10.1103/PhysRevC.61.064910}{{\em Phys. Rev.}
  {\bfseries C61} (2000) 064910},
\href{http://arxiv.org/abs/nucl-th/9812021}{{\ttfamily arXiv:nucl-th/9812021
  [nucl-th]}}.

\bibitem{Kopeliovich:2002yh}
B.~Z. Kopeliovich, J.~Nemchik, A.~Schafer, and A.~V. Tarasov, ``{Cronin effect
  in hadron production off nuclei},''
  \href{http://dx.doi.org/10.1103/PhysRevLett.88.232303}{{\em Phys. Rev. Lett.}
  {\bfseries 88} (2002) 232303},
\href{http://arxiv.org/abs/hep-ph/0201010}{{\ttfamily arXiv:hep-ph/0201010
  [hep-ph]}}.

\bibitem{Vitev:2007ve}
I.~Vitev, ``{Non-Abelian energy loss in cold nuclear matter},''
  \href{http://dx.doi.org/10.1103/PhysRevC.75.064906}{{\em Phys. Rev.}
  {\bfseries C75} (2007) 064906},
\href{http://arxiv.org/abs/hep-ph/0703002}{{\ttfamily arXiv:hep-ph/0703002
  [hep-ph]}}.

\bibitem{Aad:2012gla}
{\bfseries ATLAS} Collaboration, G.~Aad {\em et~al.}, ``{Observation of
  Associated Near-Side and Away-Side Long-Range Correlations in $\sqrt{s_{\rm
  NN}}$ = 5.02 TeV Proton-Lead Collisions with the ATLAS Detector},''
  \href{http://dx.doi.org/10.1103/PhysRevLett.110.182302}{{\em Phys. Rev.
  Lett.} {\bfseries 110} no.~18, (2013) 182302},
\href{http://arxiv.org/abs/1212.5198}{{\ttfamily arXiv:1212.5198 [hep-ex]}}.

\bibitem{ABELEV:2013wsa}
{\bfseries ALICE} Collaboration, B.~Abelev {\em et~al.}, ``{Long-range angular
  correlations of $\rm \pi$, K and p in p--Pb collisions at $\sqrt{s_{\rm NN}}$
  = 5.02 TeV},'' \href{http://dx.doi.org/10.1016/j.physletb.2013.08.024}{{\em
  Phys. Lett.} {\bfseries B726} (2013) 164--177},
\href{http://arxiv.org/abs/1307.3237}{{\ttfamily arXiv:1307.3237 [nucl-ex]}}.

\bibitem{Abelev:2012ola}
{\bfseries ALICE} Collaboration, B.~Abelev {\em et~al.}, ``{Long-range angular
  correlations on the near and away side in p--Pb collisions at $\sqrt{s_{\rm
  NN}}=5.02$ TeV},''
  \href{http://dx.doi.org/10.1016/j.physletb.2013.01.012}{{\em Phys. Lett.}
  {\bfseries B719} (2013) 29--41},
\href{http://arxiv.org/abs/1212.2001}{{\ttfamily arXiv:1212.2001 [nucl-ex]}}.

\bibitem{CMS:2012qk}
{\bfseries CMS} Collaboration, S.~Chatrchyan {\em et~al.}, ``{Observation of
  long-range near-side angular correlations in proton-lead collisions at the
  LHC},'' \href{http://dx.doi.org/10.1016/j.physletb.2012.11.025}{{\em Phys.
  Lett.} {\bfseries B718} (2013) 795--814},
\href{http://arxiv.org/abs/1210.5482}{{\ttfamily arXiv:1210.5482 [nucl-ex]}}.

\bibitem{Adam:2015bka}
{\bfseries ALICE} Collaboration, J.~Adam {\em et~al.}, ``{Forward-central
  two-particle correlations in p-Pb collisions at $\sqrt{s_{\rm NN}}$ = 5.02
  TeV},'' \href{http://dx.doi.org/10.1016/j.physletb.2015.12.010}{{\em Phys.
  Lett.} {\bfseries B753} (2016) 126--139},
\href{http://arxiv.org/abs/1506.08032}{{\ttfamily arXiv:1506.08032 [nucl-ex]}}.

\bibitem{Adler:2006xd}
{\bfseries PHENIX} Collaboration, S.~S. Adler {\em et~al.}, ``{Nuclear effects
  on hadron production in d-Au and p + p collisions at $\sqrt {s_{\rm NN}}$ =
  200 GeV},'' \href{http://dx.doi.org/10.1103/PhysRevC.74.024904}{{\em Phys.
  Rev.} {\bfseries C74} (2006) 024904},
\href{http://arxiv.org/abs/nucl-ex/0603010}{{\ttfamily arXiv:nucl-ex/0603010
  [nucl-ex]}}.

\bibitem{Abelev:2013haa}
{\bfseries ALICE} Collaboration, B.~Abelev {\em et~al.}, ``{Multiplicity
  Dependence of pion, kaon, proton and lambda production in p--Pb Collisions at
  $\sqrt{s_{\rm NN}}$ = 5.02 TeV},''
  \href{http://dx.doi.org/10.1016/j.physletb.2013.11.020}{{\em Phys. Lett.}
  {\bfseries B728} (2014) 25--38},
\href{http://arxiv.org/abs/1307.6796}{{\ttfamily arXiv:1307.6796 [nucl-ex]}}.

\bibitem{Abelev:2014zpa}
{\bfseries ALICE} Collaboration, B.~Abelev {\em et~al.}, ``{Suppression of
  $\psi$(2S) production in p--Pb collisions at $\sqrt{s_{\rm NN}}$ = 5.02
  TeV},'' \href{http://dx.doi.org/10.1007/JHEP12(2014)073}{{\em JHEP}
  {\bfseries 12} (2014) 073},
\href{http://arxiv.org/abs/1405.3796}{{\ttfamily arXiv:1405.3796 [nucl-ex]}}.

\bibitem{Adare:2013ezl}
{\bfseries PHENIX} Collaboration, A.~Adare {\em et~al.}, ``{Nuclear
  Modification of $\psi^\prime, \chi_c$, and $J/\psi$ Production in d+Au
  Collisions at $\sqrt{s_{NN}}$=200 GeV},''
  \href{http://dx.doi.org/10.1103/PhysRevLett.111.202301}{{\em Phys. Rev.
  Lett.} {\bfseries 111} no.~20, (2013) 202301},
\href{http://arxiv.org/abs/1305.5516}{{\ttfamily arXiv:1305.5516 [nucl-ex]}}.

\bibitem{Adare:2012yxa}
{\bfseries PHENIX} Collaboration, A.~Adare {\em et~al.}, ``{Cold-nuclear-matter
  effects on heavy-quark production in d+Au collisions at $\sqrt{s_{NN}}$=200
  GeV},'' \href{http://dx.doi.org/10.1103/PhysRevLett.109.242301}{{\em Phys.
  Rev. Lett.} {\bfseries 109} no.~24, (2012) 242301},
\href{http://arxiv.org/abs/1208.1293}{{\ttfamily arXiv:1208.1293 [nucl-ex]}}.

\bibitem{Abelev:2006db}
{\bfseries STAR} Collaboration, B.~I. Abelev {\em et~al.}, ``{Transverse
  momentum and centrality dependence of high-$p_T$ non-photonic electron
  suppression in Au+Au collisions at $\sqrt{s_{NN}} = 200$\,GeV},''
  \href{http://dx.doi.org/10.1103/PhysRevLett.106.159902,
  10.1103/PhysRevLett.98.192301}{{\em Phys. Rev. Lett.} {\bfseries 98} (2007)
  192301}, \href{http://arxiv.org/abs/nucl-ex/0607012}{{\ttfamily
  arXiv:nucl-ex/0607012 [nucl-ex]}}.
[Erratum: Phys. Rev. Lett.106,159902(2011)].

\bibitem{Adare:2013lkk}
{\bfseries PHENIX} Collaboration, A.~Adare {\em et~al.}, ``{Cold-Nuclear-Matter
  Effects on Heavy-Quark Production at Forward and Backward Rapidity in d+Au
  Collisions at $\sqrt{s_{NN}}=200$ GeV},''
  \href{http://dx.doi.org/10.1103/PhysRevLett.112.252301}{{\em Phys. Rev.
  Lett.} {\bfseries 112} no.~25, (2014) 252301},
\href{http://arxiv.org/abs/1310.1005}{{\ttfamily arXiv:1310.1005 [nucl-ex]}}.

\bibitem{Adare:2013xlp}
{\bfseries PHENIX} Collaboration, A.~Adare {\em et~al.}, ``{Heavy-flavor
  electron-muon correlations in $p$+$p$ and $d$+Au collisions at
  $\sqrt{s_{NN}}$ = 200 GeV},''
  \href{http://dx.doi.org/10.1103/PhysRevC.89.034915}{{\em Phys. Rev.}
  {\bfseries C89} no.~3, (2014) 034915},
\href{http://arxiv.org/abs/1311.1427}{{\ttfamily arXiv:1311.1427 [nucl-ex]}}.

\bibitem{Aaij:2013zxa}
{\bfseries LHCb} Collaboration, R.~Aaij {\em et~al.}, ``{Study of $J/\psi$
  production and cold nuclear matter effects in $p$Pb collisions at
  $\sqrt{s_{\rm NN}} = 5$ TeV},''
  \href{http://dx.doi.org/10.1007/JHEP02(2014)072}{{\em JHEP} {\bfseries 02}
  (2014) 072},
\href{http://arxiv.org/abs/1308.6729}{{\ttfamily arXiv:1308.6729 [nucl-ex]}}.

\bibitem{Khachatryan:2015uja}
{\bfseries CMS} Collaboration, V.~Khachatryan {\em et~al.}, ``{Study of B Meson
  Production in p$+$Pb Collisions at $\sqrt{s_{NN}}=$ 5.02 TeV Using Exclusive
  Hadronic Decays},''
  \href{http://dx.doi.org/10.1103/PhysRevLett.116.032301}{{\em Phys. Rev.
  Lett.} {\bfseries 116} no.~3, (2016) 032301},
\href{http://arxiv.org/abs/1508.06678}{{\ttfamily arXiv:1508.06678 [nucl-ex]}}.

\bibitem{Aad:2015ddl}
{\bfseries ATLAS} Collaboration, G.~Aad {\em et~al.}, ``{Measurement of
  differential $J/\psi$ production cross sections and forward-backward ratios
  in p + Pb collisions with the ATLAS detector},''
  \href{http://dx.doi.org/10.1103/PhysRevC.92.034904}{{\em Phys. Rev.}
  {\bfseries C92} no.~3, (2015) 034904},
\href{http://arxiv.org/abs/1505.08141}{{\ttfamily arXiv:1505.08141 [hep-ex]}}.

\bibitem{Abelev:2014hha}
{\bfseries ALICE} Collaboration, B.~Abelev {\em et~al.}, ``{Measurement of
  prompt D-meson production in p--Pb collisions at $\sqrt{s_{\rm NN}}$ = 5.02
  TeV},'' \href{http://dx.doi.org/10.1103/PhysRevLett.113.232301}{{\em Phys.
  Rev. Lett.} {\bfseries 113} no.~23, (2014) 232301},
\href{http://arxiv.org/abs/1405.3452}{{\ttfamily arXiv:1405.3452 [nucl-ex]}}.

\bibitem{Adam:2015qda}
{\bfseries ALICE} Collaboration, J.~Adam {\em et~al.}, ``{Measurement of
  electrons from heavy-flavour hadron decays in p--Pb collisions at
  $\sqrt{s_{\rm NN}} = 5.02$ TeV},''
  \href{http://dx.doi.org/10.1016/j.physletb.2015.12.067}{{\em Phys. Lett.}
  {\bfseries B754} (2016) 81--93},
\href{http://arxiv.org/abs/1509.07491}{{\ttfamily arXiv:1509.07491 [nucl-ex]}}.

\bibitem{Adam:2016wyz}
{\bfseries ALICE} Collaboration, J.~Adam {\em et~al.}, ``{Measurement of
  electrons from beauty-hadron decays in p--Pb collisions at
  $\mathbf{\sqrt{s_{\rm NN}}=5.02}$ TeV and Pb--Pb collisions at
  $\mathbf{\sqrt{s_{\rm NN}}=2.76}$ TeV},''
\href{http://arxiv.org/abs/1609.03898}{{\ttfamily arXiv:1609.03898 [nucl-ex]}}.

\bibitem{Sjostrand:2014zea}
T.~Sjöstrand, S.~Ask, J.~R. Christiansen, R.~Corke, N.~Desai, P.~Ilten,
  S.~Mrenna, S.~Prestel, C.~O. Rasmussen, and P.~Z. Skands, ``{An Introduction
  to PYTHIA 8.2},'' \href{http://dx.doi.org/10.1016/j.cpc.2015.01.024}{{\em
  Comput. Phys. Commun.} {\bfseries 191} (2015) 159--177},
\href{http://arxiv.org/abs/1410.3012}{{\ttfamily arXiv:1410.3012 [hep-ph]}}.

\bibitem{Aamodt:2008zz}
{\bfseries ALICE} Collaboration, K.~Aamodt {\em et~al.}, ``{The ALICE
  experiment at the CERN LHC},''
\href{http://dx.doi.org/10.1088/1748-0221/3/08/S08002}{{\em JINST} {\bfseries
  3} (2008) S08002}.

\bibitem{Abelev:2014ffa}
{\bfseries ALICE} Collaboration, B.~Abelev {\em et~al.}, ``{Performance of the
  ALICE Experiment at the CERN LHC},''
  \href{http://dx.doi.org/10.1142/S0217751X14300440}{{\em Int. J. Mod. Phys.}
  {\bfseries A29} (2014) 1430044},
\href{http://arxiv.org/abs/1402.4476}{{\ttfamily arXiv:1402.4476 [nucl-ex]}}.

\bibitem{Alice:2016wka}
{\bfseries ALICE} Collaboration, J.~Adam {\em et~al.}, ``{W and Z boson
  production in p--Pb collisions at $\sqrt{s_{\rm NN}}$ = 5.02 TeV},''
\href{http://arxiv.org/abs/1611.03002}{{\ttfamily arXiv:1611.03002 [nucl-ex]}}.

\bibitem{Abelev:2014epa}
{\bfseries ALICE} Collaboration, B.~Abelev {\em et~al.}, ``{Measurement of
  visible cross sections in proton-lead collisions at $\sqrt{s_{\rm NN}}$ =
  5.02 TeV in van der Meer scans with the ALICE detector},''
  \href{http://dx.doi.org/10.1088/1748-0221/9/11/P11003}{{\em JINST} {\bfseries
  9} no.~11, (2014) P11003},
\href{http://arxiv.org/abs/1405.1849}{{\ttfamily arXiv:1405.1849 [nucl-ex]}}.

\bibitem{Cacciari:2012ny}
M.~Cacciari, S.~Frixione, N.~Houdeau, M.~L. Mangano, P.~Nason, and G.~Ridolfi,
  ``{Theoretical predictions for charm and bottom production at the LHC},''
  \href{http://dx.doi.org/10.1007/JHEP10(2012)137}{{\em JHEP} {\bfseries 10}
  (2012) 137},
\href{http://arxiv.org/abs/1205.6344}{{\ttfamily arXiv:1205.6344 [hep-ph]}}.

\bibitem{Brun:1994aa}
R.~Brun, F.~Bruyant, F.~Carminati, S.~Giani, M.~Maire, A.~McPherson,
  G.~Patrick, and L.~Urban, ``{GEANT Detector Description and Simulation
  Tool}.'' Cern-w5013,~cern-w-5013,~w5013,~w-5013, 1994.

\bibitem{Adam:2016dau}
{\bfseries ALICE} Collaboration, J.~Adam {\em et~al.}, ``{Multiplicity
  dependence of charged pion, kaon, and (anti)proton production at large
  transverse momentum in p--Pb collisions at $\mathbf{\sqrt{{\textit s}_{\rm
  NN}}}$ = 5.02 TeV},''
  \href{http://dx.doi.org/10.1016/j.physletb.2016.07.050}{{\em Phys. Lett.}
  {\bfseries B760} (2016) 720--735},
\href{http://arxiv.org/abs/1601.03658}{{\ttfamily arXiv:1601.03658 [nucl-ex]}}.

\bibitem{Roesler:2000he}
S.~Roesler, R.~Engel, and J.~Ranft,
  \href{http://dx.doi.org/10.1007/978-3-642-18211-2_166}{``{The Monte Carlo
  event generator DPMJET-III},''} in {\em {Advanced Monte Carlo for radiation
  physics, particle transport simulation and applications. Proceedings,
  Conference, MC2000, Lisbon, Portugal, October 23-26, 2000}}, pp.~1033--1038.
\newblock 2000.
\newblock \href{http://arxiv.org/abs/hep-ph/0012252}{{\ttfamily
  arXiv:hep-ph/0012252 [hep-ph]}}.
\newblock
\url{http://www-public.slac.stanford.edu/sciDoc/docMeta.aspx?slacPubNumber=SLAC-PUB-8740}.
\newblock

\bibitem{ALICE:2012xs}
{\bfseries ALICE} Collaboration, B.~Abelev {\em et~al.}, ``{Pseudorapidity
  density of charged particles in $p$+Pb collisions at $\sqrt{s_{\rm NN}}=5.02$
  TeV},'' \href{http://dx.doi.org/10.1103/PhysRevLett.110.032301}{{\em Phys.
  Rev. Lett.} {\bfseries 110} no.~3, (2013) 032301},
\href{http://arxiv.org/abs/1210.3615}{{\ttfamily arXiv:1210.3615 [nucl-ex]}}.

\bibitem{Xu:2012au}
R.~Xu, W.-T. Deng, and X.-N. Wang, ``{Nuclear modification of high-pT hadron
  spectra in p+A collisions at LHC},''
  \href{http://dx.doi.org/10.1103/PhysRevC.86.051901}{{\em Phys. Rev.}
  {\bfseries C86} (2012) 051901},
\href{http://arxiv.org/abs/1204.1998}{{\ttfamily arXiv:1204.1998 [nucl-th]}}.

\bibitem{Lin:2004en}
Z.-W. Lin, C.~M. Ko, B.-A. Li, B.~Zhang, and S.~Pal, ``{A Multi-phase transport
  model for relativistic heavy ion collisions},''
  \href{http://dx.doi.org/10.1103/PhysRevC.72.064901}{{\em Phys. Rev.}
  {\bfseries C72} (2005) 064901},
\href{http://arxiv.org/abs/nucl-th/0411110}{{\ttfamily arXiv:nucl-th/0411110
  [nucl-th]}}.

\bibitem{Khachatryan:2015xaa}
{\bfseries CMS} Collaboration, V.~Khachatryan {\em et~al.}, ``{Nuclear effects
  on the transverse momentum spectra of charged particles in pPb collisions at
  $\sqrt{s_{\rm NN}} = 5.02$ TeV},''
  \href{http://dx.doi.org/10.1140/epjc/s10052-015-3435-4}{{\em Eur. Phys. J.}
  {\bfseries C75} no.~5, (2015) 237},
\href{http://arxiv.org/abs/1502.05387}{{\ttfamily arXiv:1502.05387 [nucl-ex]}}.

\bibitem{Averbeck:2011ga}
R.~Averbeck, N.~Bastid, Z.~C. del Valle, P.~Crochet, A.~Dainese, and X.~Zhang,
  ``{Reference Heavy Flavour Cross Sections in pp Collisions at $\sqrt{s}$ =
  2.76 TeV, using a pQCD-Driven $\sqrt{s}$-Scaling of ALICE Measurements at
  $\sqrt{s}$ = 7 TeV},''
\href{http://arxiv.org/abs/1107.3243}{{\ttfamily arXiv:1107.3243 [hep-ph]}}.

\bibitem{Abelev:2013yxa}
{\bfseries ALICE} Collaboration, B.~Abelev {\em et~al.}, ``{$J/\psi$ production
  and nuclear effects in p--Pb collisions at $\sqrt{s_{\rm NN}}$ = 5.02 TeV},''
  \href{http://dx.doi.org/10.1007/JHEP02(2014)073}{{\em JHEP} {\bfseries 02}
  (2014) 073},
\href{http://arxiv.org/abs/1308.6726}{{\ttfamily arXiv:1308.6726 [nucl-ex]}}.

\bibitem{Abelev:2014oea}
{\bfseries ALICE} Collaboration, B.~Abelev {\em et~al.}, ``{Production of
  inclusive $\Upsilon$(1S) and $\Upsilon$(2S) in p--Pb collisions at
  $\sqrt{s_{\rm NN}} = 5.02$ TeV},''
  \href{http://dx.doi.org/10.1016/j.physletb.2014.11.041}{{\em Phys. Lett.}
  {\bfseries B740} (2015) 105--117},
\href{http://arxiv.org/abs/1410.2234}{{\ttfamily arXiv:1410.2234 [nucl-ex]}}.

\bibitem{Abelev:2012sea}
{\bfseries ALICE} Collaboration, B.~Abelev {\em et~al.}, ``{Measurement of
  inelastic, single- and double-diffraction cross sections in proton--proton
  collisions at the LHC with ALICE},''
  \href{http://dx.doi.org/10.1140/epjc/s10052-013-2456-0}{{\em Eur. Phys. J.}
  {\bfseries C73} no.~6, (2013) 2456},
\href{http://arxiv.org/abs/1208.4968}{{\ttfamily arXiv:1208.4968 [hep-ex]}}.

\bibitem{Mangano:1991jk}
M.~L. Mangano, P.~Nason, and G.~Ridolfi, ``{Heavy quark correlations in hadron
  collisions at next-to-leading order},''
\href{http://dx.doi.org/10.1016/0550-3213(92)90435-E}{{\em Nucl. Phys.}
  {\bfseries B373} (1992) 295--345}.

\bibitem{Sharma:2009hn}
R.~Sharma, I.~Vitev, and B.-W. Zhang, ``{Light-cone wave function approach to
  open heavy flavor dynamics in QCD matter},''
  \href{http://dx.doi.org/10.1103/PhysRevC.80.054902}{{\em Phys. Rev.}
  {\bfseries C80} (2009) 054902},
\href{http://arxiv.org/abs/0904.0032}{{\ttfamily arXiv:0904.0032 [hep-ph]}}.

\bibitem{Kang:2014hha}
Z.-B. Kang, I.~Vitev, E.~Wang, H.~Xing, and C.~Zhang, ``{Multiple scattering
  effects on heavy meson production in p+A collisions at backward rapidity},''
  \href{http://dx.doi.org/10.1016/j.physletb.2014.11.024}{{\em Phys. Lett.}
  {\bfseries B740} (2015) 23--29},
\href{http://arxiv.org/abs/1409.2494}{{\ttfamily arXiv:1409.2494 [hep-ph]}}.

\bibitem{Stump:2003yu}
D.~Stump, J.~Huston, J.~Pumplin, W.-K. Tung, H.~L. Lai, S.~Kuhlmann, and J.~F.
  Owens, ``{Inclusive jet production, parton distributions, and the search for
  new physics},'' \href{http://dx.doi.org/10.1088/1126-6708/2003/10/046}{{\em
  JHEP} {\bfseries 10} (2003) 046},
\href{http://arxiv.org/abs/hep-ph/0303013}{{\ttfamily arXiv:hep-ph/0303013
  [hep-ph]}}.

\bibitem{Fujii:2015lld}
H.~Fujii and K.~Watanabe, ``{Leptons from heavy-quark semileptonic decay in p A
  collisions within the CGC framework},''
  \href{http://dx.doi.org/10.1016/j.nuclphysa.2016.03.045}{{\em Nucl. Phys.}
  {\bfseries A951} (2016) 45--59},
\href{http://arxiv.org/abs/1511.07698}{{\ttfamily arXiv:1511.07698 [hep-ph]}}.

\end{thebibliography}\endgroup
